\documentclass[zpreprint,zbstdefault]{./LaTeX/zeus/zeus_paper}
\usepackage[american]{babel}
\newcommand{\ZcoosysA}{%
The ZEUS coordinate system is a right-handed Cartesian system, with the $Z$
axis pointing in the proton beam direction, referred to as the ``forward
direction'', and the $X$ axis pointing left towards the center of HERA.
The coordinate origin is at the nominal interaction point.\xspace}

\newcommand{\Zpsrap}{%
The pseudorapidity is defined as $\eta=-\ln\left(\tan\frac{\theta}{2}\right)$,
where the polar angle, $\theta$, is measured with respect to the proton beam
direction.\xspace}

\newcommand{\ZcoosysfnAeta}{\footnote{\ZcoosysA\Zpsrap}}

\newcommand{\Zdetdesc}{%
A detailed description of the ZEUS detector can be found 
elsewhere~\cite{zeus:1993:bluebook}. A brief outline of the 
components which are most relevant for this analysis is given
below.\xspace}
\newcommand{\Zctddesc}[1]{%
Charged particles are tracked in the central tracking detector (CTD)~\citeCTD,
which operates in a magnetic field of $1.43\Tesla$ provided by a thin 
superconducting coil. The CTD consists of 72~cylindrical drift chamber 
layers, organized in 9~superlayers covering the polar angle#1 region 
\mbox{$15^\circ<\theta<164^\circ$}. The transverse-momentum resolution for
full-length tracks is $\sigma(p_T)/p_T=0.0058p_T\oplus0.0065\oplus0.0014/p_T$,
with $p_T$ in $\Gev$.}

\chardef\usc=95
\chardef\til=126
\catcode`\@=11 
\DeclareRobustCommand\xdotspace{\futurelet\@let@token\@xdotspace}
\def\@xdotspace{%
  \ifx\@let@token.\else
  \ifx\@let@token\bgroup.\else
  \ifx\@let@token\egroup.\else
  \ifx\@let@token\/.\else
  \ifx\@let@token\ .\else
  \ifx\@let@token~.\else
  \ifx\@let@token!.\else
  \ifx\@let@token,.\else
  \ifx\@let@token:.\else
  \ifx\@let@token;.\else
  \ifx\@let@token?.\else
  \ifx\@let@token/.\else
  \ifx\@let@token'.\else
  \ifx\@let@token).\else
  \ifx\@let@token-.\else
  \ifx\@let@token\@xobeysp.\else
  \ifx\@let@token\space.\else
  \ifx\@let@token\@sptoken.\else
   .\space
   \fi\fi\fi\fi\fi\fi\fi\fi\fi\fi\fi\fi\fi\fi\fi\fi\fi\fi}
\catcode`\@=12 
\newcommand{\CL}[1]{$#1\%$~C.L\xdotspace}
\newcommand{\stru}[2]{%
   \relax\ifmmode\hbox{\vrule height#1 depth#2 width0pt}%
   \else\vrule height#1 depth#2 width0pt\fi}

\newcommand{\Ronum}[1]{\uppercase\expandafter{\romannumeral#1}}
\newcommand{\ronum}[1]{\expandafter{\romannumeral#1}}
\DeclareRobustCommand{\LaTeXZ}{%
  \LaTeX\kern-.05em4\kern-.1em
  {\raisebox{-0.2ex}{$\scriptstyle\text{ZEUS}$}}\xspace}

\newcommand{\eq}[1]{(\ref{eq-#1})}

\newcommand{\fig}[1]{Fig.~\ref{fig-#1}}
\newcommand{\Fig}[1]{Figure~\ref{fig-#1}}

\newcommand{\Tab}[1]{Table~\ref{tab-#1}}
\newcommand{\taband}[2]{Tables~\ref{tab-#1} and~\ref{tab-#2}}

\newcommand{\Sect}[1]{Section~\ref{sec-#1}}

\newcommand{\Sectand}[2]{Sections~\ref{sec-#1} and~\ref{sec-#2}}

\DeclareMathAlphabet{\mathbf}{OT1}{cmr}{bx}{sl}
\newcommand{\eVdist}{\kern-0.06667em}

\newcommand{\Gev}{{\text{Ge}\eVdist\text{V\/}}}

\newcommand{\gev}{{\,\text{Ge}\eVdist\text{V\/}}}

\newcommand{\pb}{\,\text{pb}}

\newcommand{\pbi}{\,\text{pb}^{-1}}

\newcommand{\Tesla}{\,\text{T}}

\newcommand{\slashfrac}[2]{%
  \raisebox{0.5ex}{\ensuremath #1}\kern-0.12em/\kern-0.08em
  \raisebox{-.8ex}{\ensuremath #2}}

\newcommand{\sqr}[3]{%
    {\vcenter{\hrule height.#3ex\hbox{\vrule width.#2ex height#1ex
     \kern#1ex\vrule width.#3ex}\hrule height.#2ex}}}

\newcommand{\widebar}[1]{%
   \mkern1.5mu\overline{\mkern-1.5mu#1\mkern-1.mu}\mkern1.mu}
\catcode`\@=11 
\newcommand{\parenbar}{\mathpalette\p@renb@r}
\def\p@renb@r#1#2{\vbox{%
  \ifx#1\scriptscriptstyle \dimen@.7em\dimen@ii.2em\else
  \ifx#1\scriptstyle \dimen@.8em\dimen@ii.25em\else
  \dimen@1em\dimen@ii.4em\fi\fi \offinterlineskip
  \ialign{\hfill##\hfill\cr
    \vbox{\hrule width\dimen@ii}\cr
    \noalign{\vskip-.3ex}%
    \hbox to\dimen@{$\mathchar300\hfil\mathchar301$}\cr
    \noalign{\vskip-.3ex}%
    $#1#2$\cr}}}
\catcode`\@=12 

\newcommand{\nubar}{\widebar{\nu}}

\newcommand{\rnge}{\hbox{$\,\text{--}\,$}}

\newcommand{\tilS}{\tilde S}

\newcommand{\IP}{{\rm I$\kern-0.01667em$P}\xspace}

\newcommand{\LQ}{{\rm LQ}}

\mathchardef\qsm=63
\mathchardef\pls=43
\mathchardef\mns=512
\mathchardef\plm=518
\mathchardef\eql=61
\mathchardef\smallleft=300
\mathchardef\smallright=301
\mathchardef\les=316
\mathchardef\gre=318
\mathchardef\leq=532
\mathchardef\grq=533

\catcode`\@=11 
\newcounter{pict@width}
\newcounter{pict@height}
\newlength{\pict@scale}
\newcommand{\psfigadd}[4]{%
\setcounter{pict@width}{1*\ratio{#2+\pict@scale/2}{\pict@scale}}
\setcounter{pict@height}{1*\ratio{#3+\pict@scale/2}{\pict@scale}}
\setlength{\unitlength}{\pict@scale}
\hbox to #2{\hspace{-\fill}\begin{picture}(\thepict@width,\thepict@height)
\put(0,0){\psfig{figure=#1,width=#2,height=#3,clip=}}
\SetScale{0.283466457}
\SetWidth{1.763889}
{#4}
\end{picture}}
}
\newcounter{pict@widthfst}
\newcounter{pict@widthscd}
\newcounter{pict@widthtot}
\newcommand{\psfigaddtwo}[7]{%
\setcounter{pict@widthfst}{1*\ratio{#2+\pict@scale/2}{\pict@scale}}
\setcounter{pict@widthscd}{1*\ratio{#2+#4+\pict@scale/2}{\pict@scale}}
\setcounter{pict@widthtot}{1*\ratio{#2+#4+#6+\pict@scale/2}{\pict@scale}}
\setcounter{pict@height}{1*\ratio{#3+\pict@scale/2}{\pict@scale}}
\setlength{\unitlength}{\pict@scale}
\hbox{\hspace{-\fill}\begin{picture}(\thepict@widthtot,\thepict@height)
\put(0,0){\psfig{figure=#1,width=#2,height=#3,clip=}}
\put(\thepict@widthscd,0){\psfig{figure=#5,width=#6,height=#3,clip=}}
\SetScale{0.283466457}
\SetWidth{1.763889}
{#7}
\end{picture}}
}
\newcommand{\psfigror}[4]{%
\setcounter{pict@width}{1*\ratio{#2+\pict@scale/2}{\pict@scale}}
\setcounter{pict@height}{1*\ratio{#3+\pict@scale/2}{\pict@scale}}
\setlength{\unitlength}{\pict@scale}
\hbox{\begin{picture}(\thepict@width,\thepict@height)
\put(0,\thepict@height){\psfig{figure=#1,width=#3,height=#2,clip=,angle=270}}
\SetScale{0.283466457}
\SetWidth{1.763889}
{#4}
\end{picture}}
}
\newcommand{\psfigrol}[4]{%
\setcounter{pict@width}{1*\ratio{#2+\pict@scale/2}{\pict@scale}}
\setcounter{pict@height}{1*\ratio{#3+\pict@scale/2}{\pict@scale}}
\setlength{\unitlength}{\pict@scale}
\hbox{\begin{picture}(\thepict@width,\thepict@height)
\put(0,0){\psfig{figure=#1,width=#3,height=#2,clip=,angle=90}}
\SetScale{0.283466457}
\SetWidth{1.763889}
{#4}
\end{picture}}
}
\catcode`\@=12 
\newlength\listtextwidth

\newcommand{\pcite}[1]{{\protect\cite{#1}}}

\catcode`\@=11 
\newlength{\@tabfninsert}
\newlength{\@tabfnwidth}
\newcommand{\tabfootnote}[2]{%
  \setlength{\@tabfninsert}{0.8em}
  \setlength{\@tabfnwidth}{\textwidth}
  \addtolength{\@tabfnwidth}{-\@tabfninsert}
  \addtolength{\@tabfnwidth}{-0.4em}
  \noindent\makebox[\@tabfninsert][r]{\footnotesize$^{#1}$\hfil}\hfill%
  \parbox[t]{\@tabfnwidth}{\footnotesize #2\hfill}}
\catcode`\@=12 

\newcommand{\br}[1]{{\beta_{#1}}}
\newcommand{\MLQ}{M_{\LQ}}
\newcommand{\sigNWA}{\sigma^{\textrm{\it NWA}}}
\newcommand {\ptmiss}{\mbox{$\not\hspace{-0.55ex}{P}_t$}}
\newcommand{\Et}{{E_t}}
\newcommand{\Etjet}{{E_t^{\rm jet}}}
\newcommand{\Ettaujet}{{E_t^{\tau-\rm jet}}}
\newcommand{\phimiss}{{\phi_{\rm miss}}}

\newcommand{\pttrk}{{P_t^{\rm trk}}}
\newcommand{\mjet}{{M_{\rm jet}}}

\newcommand{\femc}{{f_{\rm EMC}}}
\newcommand{\flt}{{f_{\rm LT}}}
\newcommand{\rp}{$\not$\kern-0.pt$R_p$}

\newcommand{\qone}{{q_{_1}}}
\newcommand{\qi}{q_i}
\newcommand{\qf}{q_f}
\newcommand{\qflow}{q}
\newcommand{\qe}{{q_\alpha}}
\newcommand{\qeb}{{{\bar q}_\alpha}}
\newcommand{\ql}{{q_\beta}}
\newcommand{\qlb}{{{\bar q}_\beta}}
\newcommand{\ShL}{S_{1/2}^L}
\newcommand{\tShL}{\tilde{S}_{1/2}^L}
\newcommand{\ShR}{S_{1/2}^R}
\newcommand{\SzL}{S_0^L}
\newcommand{\SoL}{S_1^L}
\newcommand{\SzR}{S_0^R}
\newcommand{\tSzR}{\tilde{S}_0^R}
\newcommand{\VzL}{V_0^L}
\newcommand{\VhL}{V_{1/2}^L}
\newcommand{\VoL}{V_1^L}
\newcommand{\VzR}{V_0^R}
\newcommand{\tVzR}{\tilde{V}_0^R}
\newcommand{\tVhL}{\tilde{V}_{1/2}^L}
\newcommand{\VhR}{V_{1/2}^R}
\newcommand{\figs}[1]{Figs.~\ref{fig-#1}}
\newcommand{\Figs}[1]{Figures~\ref{fig-#1}}

\def\citeCTD{{\cite{%
nim:a279:290,*npps:b32:181,*nim:a338:254%
}}\xspace}
\def\citeCAL{{\cite{%
nim:a309:77,*nim:a309:101,*nim:a321:356,*nim:a336:23%
}}\xspace}

\begin{document}
\title{Search for lepton-flavor violation\\
in {\boldmath $e^+p$} collisions at HERA
}                                                       


\author{ZEUS Collaboration}
\draftversion{1.8}
\date{\today}

\abstract{
A search has been made for lepton-flavor-violating interactions of the
type $e^+p\to \ell X$, where $\ell$ denotes a $\mu$ or $\tau$ with
high transverse momentum, at a
center-of-mass energy, $\sqrt{s}$, of $300\gev$ with an
integrated luminosity of $47.7\pbi$ using the ZEUS detector at HERA.
No evidence was found for lepton-flavor violation 
and constraints were derived on leptoquarks (LQs) that
could mediate such interactions. For LQ masses below $\sqrt{s}$,
limits are set on $\lambda_{e\qone}\sqrt{\br{\ell\qflow}}$,
where $\lambda_{e\qone}$ is the coupling of the LQ to an electron and a
first-generation quark $\qone$ and $\br{\ell\qflow}$ is the branching
ratio of the LQ to  $\ell$ and a quark. For LQ masses
exceeding $\sqrt{s}$, limits are set on the four-fermion contact-interaction
term $\lambda_{e\qe}\lambda_{\ell\ql}/M_{{\rm LQ}}^2$
for leptoquarks that couple to an electron and a quark $\qe$ and
also to $\ell$ and a quark $\ql$. Some of the limits are also
applicable to lepton-flavor-violating processes mediated by squarks 
in $R$-parity-violating
supersymmetric models. In some cases involving heavy quarks and
especially for $\ell=\tau$, the ZEUS limits are the most stringent
published to date.
}

\prepnum{DESY-01-222}

\makezeustitle

\pagenumbering{Roman}                                                                              
                                                   %
\begin{center}                                                                                     
{                      \Large  The ZEUS Collaboration              }                               
\end{center}                                                                                       
  S.~Chekanov,                                                                                     
  M.~Derrick,                                                                                      
  D.~Krakauer,                                                                                     
  S.~Magill,                                                                                       
  B.~Musgrave,                                                                                     
  A.~Pellegrino,                                                                                   
  J.~Repond,                                                                                       
  R.~Yoshida\\                                                                                     
 {\it Argonne National Laboratory, Argonne, Illinois 60439-4815}                            
\par \filbreak                                                                                     
  M.C.K.~Mattingly \\                                                                              
 {\it Andrews University, Berrien Springs, Michigan 49104-0380}                                    
\par \filbreak                                                                                     
  P.~Antonioli,                                                                                    
  G.~Bari,                                                                                         
  M.~Basile,                                                                                       
  L.~Bellagamba,                                                                                   
  D.~Boscherini,                                                                                   
  A.~Bruni,                                                                                        
  G.~Bruni,                                                                                        
  G.~Cara~Romeo,                                                                                   
  L.~Cifarelli,                                                                                    
  F.~Cindolo,                                                                                      
  A.~Contin,                                                                                       
  M.~Corradi,                                                                                      
  S.~De~Pasquale,                                                                                  
  P.~Giusti,                                                                                       
  G.~Iacobucci,                                                                                    
  G.~Levi,                                                                                         
  A.~Margotti,                                                                                     
  T.~Massam,                                                                                       
  R.~Nania,                                                                                        
  F.~Palmonari,                                                                                    
  A.~Pesci,                                                                                        
  G.~Sartorelli,                                                                                   
  A.~Zichichi  \\                                                                                  
  {\it University and INFN Bologna, Bologna, Italy}
% ...
\newpage                                                                                           
% ...
                                                           %
                                                           %
\newpage   
\selectlanguage{american}
\pagenumbering{arabic} 
\pagestyle{plain}
\section{Introduction}\label{sec-int}

In the Standard Model (SM), lepton flavor is conserved.
While the reported observation of neutrino
oscillations\cite{prl:81:1562,prl:87:071301}
implies that lepton-flavor violation (LFV) does occur,
minimal extensions to the SM\cite{ptp:28:870,*arevns:49:481}
that allow for finite neutrino masses and thereby account for neutrino oscillations
do not predict detectable rates of LFV 
at current collider experiments. However, many extensions of the SM, such as
grand unified theories~\cite{pr:d10:275,*prl:32:438,*prep:72:185},
models based on supersymmetry~\cite{prep:110:1,*prep:117:75},
compositeness~\cite{pl:b153:101,*pl:b167:337} or
technicolor~\cite{np:b155:237,*np:b168:69,*pr:d20:3404,*prep:74:277}
involve LFV interactions at fundamental levels. 

In high-energy positron--proton collisions at HERA, 
reactions of the type $e\qi\to\ell\qf$, where $\qi$ and $\qf$
denote initial- and
final-state quarks and $\ell$ denotes a $\mu$ or a $\tau$ with high
transverse momentum, can be detected
with high efficiency and small background.
Indirect searches for such reactions have yielded very strong
constraints~\cite{zfp:c61:613}
for cases where $\qi$ and $\qf$ are light quarks. However,
in some cases involving
heavy quarks, especially when $\ell=\tau$, the sensitivity of HERA extends beyond
existing low-energy limits.

This paper reports on a search for LFV processes in $e^+p$ collisions using data
collected by the ZEUS experiment from 1994 to 1997
with an integrated luminosity, $\cal L$, of $47.7\pbi$. Previous
searches for LFV at HERA have been reported by
ZEUS~\cite{zfp:c73:613}~( ${\cal L} \sim 4 \pbi$) and
H1~\cite{epj:c11:447}~(${\cal L} \sim 37 \pbi$).


\section{Phenomenology}
There are several mechanisms whereby lepton flavor can be violated in $ep$
collisions. This paper considers two main possibilities: leptoquarks and
R-parity-violating squarks.
\subsection{Leptoquarks}\label{sec-theory}
Leptoquarks (LQs) are bosons that carry both lepton (L) and baryon (B)
numbers and have lepton-quark Yukawa couplings. Such bosons arise
naturally in unified theories that arrange quarks and
leptons in common multiplets. A LQ that couples to leptons of two
different generations would induce LFV.
The Buchm\"uller-R\"uckl-Wyler (BRW) model~\cite{pl:b191:442}, which assumes the most general Lagrangian
with ${\rm SU(3)_C \times SU(2)_L \times U(1)_Y}$ invariant couplings
of a LQ to a lepton and a quark,
is used to classify LQ species and to calculate cross sections for
LQ-mediated processes.
The following additional assumptions were made to simplify the models
under consideration:
\begin{enumerate}
\item one LQ species dominates the cross section of the process;
\item members of each ${\rm SU(2)}$ multiplet are degenerate in mass;
\item LQs couple to either left-handed or right-handed leptons, but not both.
\end{enumerate}

There are 10 different LQ states in the BRW model, 
four of which can couple to both left- and right-handed leptons.
Because of the third assumption above, models in which these
states have left- or right-handed couplings will be treated separately in this
analysis.
Each state is characterized by spin $J=0$ or $1$, weak isospin $T=0,1/2$ or $1$ and
fermion number $F=0$ or $\pm2$ (where $F=3B+L$). 
Following the Aachen notation~\cite{zfp:c46:679}, scalar
($J=0$) and vector ($J=1$) LQs are denoted $S^{\chi}_T$ and $V^{\chi}_T$,
respectively, where $\chi=L,R$ denotes the chirality of the lepton that
couples to the LQ.
When two different hypercharge states are allowed, one is distinguished
by a tilde.
In this paper, LQs with couplings $\lambda_{e\qe}$
to an electron and a quark $\qe$, and  $\lambda_{\ell\ql}$ to
a lepton $\ell$ ($\mu$ or $\tau$) and a quark $\ql$, are considered\footnote{
Note that in the BRW model, some $\chi=L$ LQs also have neutrino-quark 
couplings; these
couplings are fixed by ${\rm SU(2)_L}$ invariance to be equal to
the corresponding charged lepton-quark couplings.}. 
The subscripts $\alpha$ and $\beta$ label the quark generations.
The LQ species determines whether $\qe$ or $\ql$ are up- or down-type
quarks.
In addition to mediating LFV interactions, such LQs would also mediate
flavor-conserving interactions with an  $e$ or a $\nu_e$ in the final state.
These final states were not searched for in this analysis, but they were taken
into account in calculating branching ratios (the same is true for final
states with $\nu_\mu$ or $\nu_\tau$).

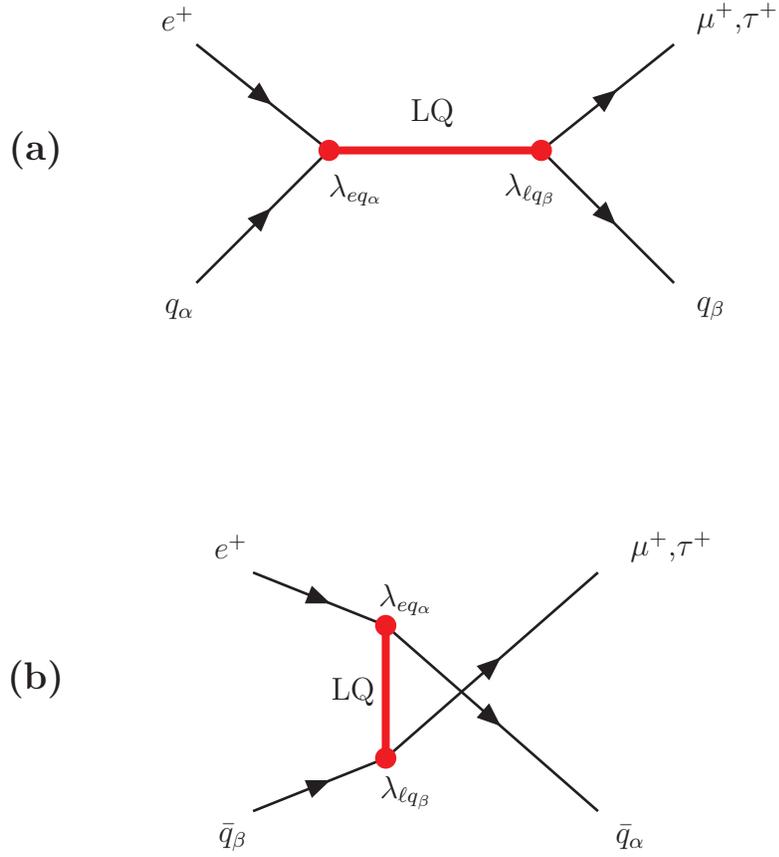
\begin{figure}[t]
\vspace*{5.cm}
\begin{picture}(150.,100.)(0.,-400.)
{
\SetOffset(-60.,-100.)
\SetScale{2}
\Text(130.,10.)[cc]{\large \bf (a)}
\Text(190.,60)[rc]{$e^+$}
\ArrowLine(95.,25)(120.,5.)
\ArrowLine(95.,-20)(120.,5.)
\Text(190.,-50)[rc]{$\qe$}
\SetColor{Red}
\Vertex(120.,5.){2.}
\Text(251,-5)[cc]{$\lambda_{e \qe}$}
\SetColor{Red}
\SetWidth{1.5}
\Line(120.,5.)(160.,5.)
\Text(280,24)[cc]{LQ}
\Text(317,-5.)[cc]{$\lambda_{\ell\ql}$}
\SetWidth{0.5}
\SetColor{Black}
\ArrowLine(160.,5.)(185.,25.)
\Text(380,60.)[lc]{$\mu^+$,$\tau^+$}
\ArrowLine(160.,5.)(185.,-20.)
\SetColor{Red}
\Vertex(160.,5.){2.}
\SetColor{Black}
\Text(380,-50.)[lc]{$\ql$}}

{
\SetOffset(-430.,-300.)
\SetScale{2}
\Text(500.,10.)[cc]{\large \bf (b)}
\Text(580.,60.)[rc]{$e^+$}
\ArrowLine(290.,25.)(315.,15.)
\ArrowLine(315.,15.)(355.,-20.)
\Text(725.,-50.)[cc]{$\qeb$}
\SetColor{Red}
\Vertex(315.,15.){2.}
\Text(640.,40)[cc]{$\lambda_{e \qe}$}
\SetWidth{1.5}
\Line(315.,15.)(315.,-10.)
\Text(620.,5)[cc]{LQ}
\Text(640.,-33.)[cc]{$\lambda_{\ell\ql}$}
\SetWidth{0.5}
\SetColor{Black}
\Text(580.,-50)[rc]{$\qlb$}
\ArrowLine(290.,-20.)(315.,-10.)
\ArrowLine(315.,-10.)(355.,25.)
\SetColor{Red}
\Vertex(315.,-10.){2.}
\SetColor{Black}
\Text(725,60.)[lc]{$\mu^+$,$\tau^+$}
}
\end{picture}
\caption{\it (a) $s$-channel and (b) $u$-channel diagrams
  contributing to LFV processes induced by $F=0$ LQs. In $e^+p$
  scattering, $|F|=2$ LQs couple to antiquarks in the $s$-channel and to
  quarks in the $u$-channel.} 
\label{fig-LQFEY}
\end{figure}

In $ep$ collisions, if the LQ mass, $\MLQ$, is below
$\sqrt{s}$ (low-mass LQs), the LQ is predominantly produced as an
$s$-channel resonance, as shown in \fig{LQFEY}(a).
In this case, only incident $u$ or $d$ quarks,
denoted $q_1$, which couple to the incident positron to produce
$F=0$ LQs, are considered. In the $e^+p$ data analyzed here,
the production cross section for $F=0$ LQs is much larger than for
$F=-2$ LQs, assuming that $\MLQ$ is sufficiently large so that the 
production is valence-quark dominated, since $F=-2$ LQs
would be produced via $e^+\bar{q}$ fusion.

For small values of the Yukawa coupling, $\lambda_{eq_1}$,
the resonance width becomes negligible and the $s$-channel
Breit-Wigner line shape can be approximated (neglecting radiative
effects) by a $\delta$-function at $\MLQ=\sqrt{xs}$, where $x$ is
the Bjorken variable in deep inelastic scattering (DIS). This leads to
the Narrow Width Approximation (NWA)
\begin{equation}
\sigma^{\textrm{\it NWA}}_{T_3}(s,\MLQ)=\left(J+1\right)\frac{\pi
  \lambda_{e\qone}^2}{4s}
C_{T_3}\,q_{1}\left(x=\frac{\MLQ^2}{s},Q^2_0\right),
\label{eq-lowmas}
\end{equation}
where $T_3$ is the third component of the weak isospin, $C_{T_3}$ is the square of 
the relevant ${\rm SU(2)}$
Clebsch-Gordan coefficient, and $q_1(x,Q^2_0)$ is the valence-quark
density in the proton evaluated at the scale $Q^2_0=\MLQ^2$.
The total production 
cross section for a given LQ is given by the sum over all states of 
the ${\rm SU(2)}$ multiplet that couple to a positron and a quark.  
The NWA becomes inaccurate if $q_1(x,Q_0^2)$ varies significantly with $x$
on a scale corresponding to the LQ width,
$\Gamma_{LQ}\propto\lambda^2\MLQ$.
In the present analysis, this occurs only when $\MLQ$ is
close to $\sqrt{s}$ ($x\rightarrow 1$). In this region,
$q_1$ falls steeply with $x$ and the
convolution of $q_1$ with the Breit-Wigner line shape results in
contributions to the cross section from quarks with $x$ below
the resonant peak.
These non-resonant contributions to the cross section are neglected in the NWA.
For LQs that couple to $u$ ($d$) quarks, with $\MLQ=270\gev$ ($250\gev$)
and $\lambda_{e\qone}=0.3$, $\sigNWA$ underestimates the cross section
by $\simeq 20\%$.
The rate for LFV events is proportional to $\sigma\br{\ell\qflow}$ where
$\br{\ell\qflow}$ is the branching ratio to the $\ell\qflow$ final state.
In this paper, the NWA is used to calculate cross sections for $\MLQ<\sqrt{s}$,
so that limits on $\sigma\br{\ell\qflow}$ can be simply converted to limits on
$\lambda_{e\qone}\sqrt{\br{\ell\qflow}}$.
The NWA underestimates the cross
section, leading to conservative limits.

If $\MLQ>\sqrt{s}$ (high-mass LQs), both
$s$- and $u$-channel diagrams contribute, see \fig{LQFEY}.
If $\MLQ \gg \sqrt{s}$, the LQ propagator contracts to
a four--fermion contact interaction and the cross
section is proportional to $[\lambda_{e\qe}\lambda_{\ell\ql}/M_\LQ^2]^2$.
In this high-mass approximation (HMA), the cross section for an $F=0$ LQ, in
$e^+p$ collisions, can be written as
\begin{equation}
\sigma^{\textrm{\it HMA}}_{F=0} = \frac{s}{32\pi} \left[\frac{\lambda_{e\qe}\lambda_{\ell\ql}}
{\MLQ^2}\right]^2 \left[\int dx\,dy\,x\,\qe(x,\hat{s})\,f(y) 
+ \int dx\,dy\,x\,\qlb(x,-\hat{u})\,g(y)\right], 
\label{eq-highmas}
\end{equation}
with
\begin{displaymath}
\begin{array}{c c c}
f(y)=
\left\{\begin{array}{l l}  
1/2 & \textrm{scalar~LQ}\\
2(1-y)^2 & \textrm{vector~LQ}
\end{array}\right.
;
& &
g(y)=
\left\{\begin{array}{l l}  
(1-y)^2/2 & \textrm{scalar~LQ}\\
2 & \textrm{vector~LQ}
\end{array}\right.
,
\end{array}
\end{displaymath}
where $y$ is the inelasticity, $\hat{s}=sx$ and $\hat{u}=sx(y-1)$ are the
scales at which the quark densities $\qe$ and $\qlb$ are
evaluated.
The first and second integrals in \eq{highmas} are
due to the $s-$  and $u-$channel contributions, respectively ($|F|=2$ LQs couple
a quark in the $u$-channel and an anti-quark in the $s-$channel).
The accuracy of the HMA increases with increasing LQ mass.
For $\MLQ > 600\gev$, the minimum mass considered for this high-mass analysis,
the accuracy is better than $10\%$.

In the high-mass case,  LQ
scenarios are characterized by the 14 LQ species,
the three generations of $\qe$ and $\ql$, and the two possible final-state leptons,
leading to a total of 252 different LQ scenarios.

NLO QCD corrections~\cite{zfp:c74:611,zfp:c75:453} were applied only to the
NWA production cross section for scalar LQs, since no calculation is available
for vector LQs or for high-mass scalar LQs.
These corrections increase the production cross section by $\simeq
15\%$ at $\MLQ=150\gev$, increasing to $\simeq 30\%$ at
$\MLQ=250\gev$.

Corrections for QED initial-state radiation (ISR),
evaluated using the Weizs\"acker-Williams
approximation~\cite{zp:88:612,pr:45:729}, were applied to both the low-
and high-mass cases. The
QED ISR correction reduces the NWA cross section by $\sim 3\%$ at
$\MLQ=150\gev$ and by $\sim 25\%$ when $\MLQ$ approaches the
kinematic limit.
For high-mass LQs, QED ISR corrections, evaluated at $\MLQ=600\gev$, 
were applied. They lower the cross section by less than $5\%$;
the corrections decrease at higher masses.

\subsection{{\boldmath $R$}-parity-violating squarks}\label{sec-susy}

Supersymmetry (SUSY), which links bosons and fermions, is a promising
extension to the SM. It assumes a supersymmetric partner for each SM
particle, a bosonic partner for a fermion and vice-versa. 
$R$-parity is a multiplicative quantum number defined as
$R_p=(-1)^{3B+L+2J}$. 
For SM particles, $R_p=1$; for SUSY particles (sparticles), $R_p=-1$. In
$R_p$-conserving processes, sparticles are pair produced and the
lightest supersymmetric particle (LSP) is stable. In models with
$R_p$ violation ({\rp}), single SUSY-particle production is
possible and the LSP decays into SM particles. Of special
interest for HERA are \rp\, Yukawa couplings that couple a squark (SUSY
partner of a quark) to a lepton and a quark, which are described in 
the superpotential by the term~\cite{pr:d40:2987}
$\lambda^\prime_{ijk}L^iQ^j\overline{D}^k$, where $i$, $j$ and
$k$ are generation indices, $L$ and $Q$ denote the left-handed
lepton and quark-doublet superfields, respectively and $\overline{D}$ denotes the
right-handed quark-singlet chiral superfield.
Expansion of the superfields using four-component Dirac notation yields
\begin{equation}
{\cal L} = \lambda'_{ijk}\left[
 - {\tilde e}^i_L  u^j_L {\overline d}^k_R
 - e^i_L {\tilde u}^j_L {\overline d}^k_R 
 - ({\overline e}^i_L)^c  u^j_L ({\tilde d}^k_R)^*
 +  {\tilde\nu}^i_L  d^j_L{\overline d}^k_R
 +  \nu^i_L {\tilde d}^j_L  {\overline d}^k_R
 + ({\overline\nu}^i_L)^c d^j_L ({\tilde d}^k_R)^* \right]
 + {\mbox{\rm h.~c.}}
\end{equation}
The superscript $c$ denotes charge conjugation and the
asterisk denotes complex conjugation of scalar fields. For $i=1$, the second
and third terms will result in $\tilde{u}^j$ and $\tilde{d}^k$ production in
$ep$ collisions. Identical terms appear in the
Lagrangians for the scalar leptoquarks $\tilS^L_{1/2}$ and $S^L_0$~\cite{np:b397:3}.
The coupling $\lambda_{1j1}^\prime$
gives rise to the reaction $e^+d\to\tilde{u}_L^j$,
while the coupling $\lambda_{11k}^\prime$
would cause the reaction  $e^-u\to\tilde{d}_R^k$.

Lepton-flavor violation would occur in models with two non-zero Yukawa
couplings involving different lepton generations. For example,
non-zero values of $\lambda_{1j1}^\prime$ and $\lambda_{ijk}^\prime$ ($i=2,3$)
would yield the process $e^+d \to \tilde{u}^j \to\ell^+d^k$, where
$i=2,3$ corresponds to $\ell=\mu,\tau$.  Squarks also undergo
$R_p$-conserving decays to a quark and a gaugino, which were not
considered in this analysis.  Low-mass coupling limits on $\tilde{S}_{1/2}^L$
LQs can be interpreted, using 
$\lambda_{e \qone}\sqrt{\br{\ell\qflow}}
         =\lambda_{1j1}^\prime \sqrt{\br{\tilde u^j\to\ell\qflow}}$,
as limits on $\tilde{u}^j$ squarks
that couple to $e\qone$ and to $\ell\qflow$.
In the low-mass case, the limits apply for any final-state quark $q$
(except top). High-mass LQ limits can also be applied to squarks as
described in \Sect{himassLQ}.

\section{The ZEUS detector}
\label{sec-exp}

\Zdetdesc
\Zctddesc{\ZcoosysfnAeta} The CTD was used to reconstruct tracks of
isolated muons and charged $\tau$-decay products. It was also used to
determine the interaction vertex with a typical resolution of 4 mm (1
mm) along (transverse to) the beam direction.

The high-resolution uranium--scintillator calorimeter (CAL)~\citeCAL consists 
of three parts: the forward (FCAL), the barrel (BCAL) and the rear (RCAL)
calorimeters.
The calorimeters are
subdivided into towers each of which subtends a solid angle from $0.006$ to $0.04$
steradians. Each tower is longitudinally segmented into an electromagnetic
(EMC) section and two hadronic (HAC) sections (one in RCAL). Each HAC section
consists of a single cell, while the EMC
section of each tower is further subdivided transversely
into four cells (two in RCAL).
The CAL energy resolutions, as measured under
test-beam conditions, are $\sigma(E)/E=0.18/\sqrt{E}$ for electrons and
$\sigma(E)/E=0.35/\sqrt{E}$ for hadrons ($E$ in $\Gev$). The arrival time of CAL energy deposits is measured with
sub-nanosecond resolution for energy deposits above $4.5\gev$, 
allowing the rejection of non-$ep$  background.

The FMUON detector~\cite{zeus:1993:bluebook} consists of layers of limited streamer
tubes and drift-chamber planes located up to 10\,m from the
interaction point. The toroidal
magnetic fields of the iron yoke (1.4 T) that surrounds the
CAL and of two toroids (1.6 T) located about 9\,m from the
interaction point enable muon momentum measurements to be made.
The FMUON tags high-momentum muons (muons with momenta below $5\gev$
are unlikely to emerge from the FCAL) with
polar angles in the range $8^\circ< \theta <20^\circ$, extending
well beyond the CTD acceptance. 

The luminosity was measured by the luminosity detector (LUMI) from the
rate of the Bethe-Heitler process $e^+p \rightarrow e^+\gamma
p$~\cite{desy-92-066,*zfp:c63:391,*desy-01-041}, where the photon is detected in a
lead--scintillator calorimeter located at $Z=-107$\,m in the HERA
tunnel. The uncertainty on the luminosity measurement was 1.6\%. 


\section{Monte Carlo simulation}
\label{sec-MC}
The simulation of the LQ signal, including both $s$- and $u$-channel processes,
was performed using the generators
LQMGEN 1.0~\cite{thesis:silverstein:1996}
(low-mass LQs) and LQGENEP 1.0~\cite{cpc:141:83}
(high-mass LQs) based on the BRW model~\cite{pl:b191:442}. Both generators are interfaced to JETSET 7.4~\cite{cpc:82:74}
to simulate hadronization and particle decays. 

The following SM backgrounds were considered:
charged current (CC) and neutral current (NC) deep inelastic scattering (DIS)
were simulated using
DJANGO6 2.4~\cite{cpc:81:381}, with the color--dipole model ARIADNE 4.08~\cite{cpc:71:15} 
used to simulate the hadronic final state.
Elastic and inelastic $\gamma\gamma\to\ell^+\ell^-$ reactions
were simulated with LPAIR~\cite{np:b229:347}.
EPVEC 1.0~\cite{np:b375:3} was used to simulate $W$ production.
Photoproduction processes were simulated with HERWIG 5.8~\cite{cpc:67:465}. 
The ZEUS detector and trigger were simulated with a program based on
GEANT 3.13~\cite{tech:cern-dd-ee-84-1}. The simulated events were
processed by the same reconstruction programs as the data.


\section{Kinematic quantities}
Global calorimeter sums were calculated as follows:
each calorimeter cell $i$ with an energy deposit $E_i$ above a
threshold was assigned a four-momentum $P^\mu$, defined as
$P^\mu=
(E_i,E_i\cos\phi_i\sin\theta_i,E_i\sin\phi_i\sin\theta_i,E_i\cos\theta_i)$,
where $\phi_i$ and $\theta_i$
are the azimuthal and polar angles of the cell center relative to the event
vertex.
The total four-momentum deposited in the calorimeter $(E,P_X,P_Y,P_Z)$ is
given by the sum of the four-momenta for all cells. The transverse
energy, $\Et$, is given by $\sum_i E_i\sin\theta_i$. The missing transverse
momentum, $\ptmiss$, is given by $\sqrt{P_X^2+P_Y^2}$.
The azimuth assigned to $\ptmiss$, $\phimiss$, was defined by
$\cos\phimiss=-P_X/\ptmiss$ and $\sin\phimiss=-P_Y/\ptmiss$.
Jets used in identifying hadronic $\tau$ decays were reconstructed using
an $(\eta,\phi)$ cone algorithm~\cite{epj:c2:61} with cone radius $R=1$. The inputs to the jet
algorithm were the four-momentum vectors of each calorimeter cell.
The invariant mass of a jet, $\mjet$, was calculated from the sum of all
four-momentum vectors assigned to the jet. The transverse energy of a
jet was denoted by $\Etjet$.

The $E-P_Z$ of the initial state is twice the positron beam energy,
$2E_e=55\gev$. For events that are fully contained in the calorimeter
(ignoring particles escaping through the forward beam hole, which carry
negligible $E-P_Z$), the measured $E-P_Z$ should be near $2E_e$. In
photoproduction processes, where the final-state positron escapes through
the rear beam hole, the $E-P_Z$ spectrum falls steeply, so that a cut
on $E-P_Z$ is useful in reducing such backgrounds.
Events with high-energy muons, which deposit only a small fraction of their
energy in the calorimeter, will also have $E-P_Z$ substantially below $2E_e$.
In the search for the $e\to\mu$ transition
(see \Sectand{MuEventSel}{MuonId}), a cut was made on the
quantity $E-P_Z+\Delta_\mu$, where
 $\Delta_\mu=\ptmiss(1-\cos\theta_\mu)/\sin\theta_\mu$ is an estimate
of the $E-P_Z$ carried by the muon, assuming that the transverse
momentum of the muon is equal to $\ptmiss$;
$\theta_\mu$ is the polar angle of the muon track.

\section{Event selection for the {\boldmath $e\to\mu$} transition}
Events from the reaction $ep \to
\mu X$, mediated by a heavy LQ, would be characterized by a high-transverse-momentum ($P_t$)
muon balanced by a jet.
Since only a small fraction of the muon energy is deposited in the
calorimeter, these events would have a large $\ptmiss$.
The offline event selection consisted of two steps: a pre-selection of
events with $\ptmiss$ and a final selection requiring an identified
muon.

\subsection{Pre-selection}
\label{sec-MuEventSel}

The trigger, which is identical to that used in the
CC DIS measurement described elsewhere~\cite{epj:c12:411},
was based on a cut on $\ptmiss$ with a considerably lower
threshold than the selection cuts described below.
After applying cuts to reject non-$ep$ backgrounds
(mainly cosmic rays and beam-gas interactions), the following pre-selection 
requirements were imposed:
\begin{itemize}
\item a reconstructed vertex with $Z$ coordinate $|Z_{\textrm{VTX}}| < 50$\,cm;
\item $ \ptmiss > 15\gev$ and $ \ptmiss/\sqrt{\Et} > 2.5 \sqrt{\Gev}$;
\item $25\gev < E-P_Z+\Delta_\mu< 100\gev$; 
\item no electron with energy larger than $10 \gev$.
\end{itemize}

\begin{figure}[htb]
\begin{center}
\epsfig{figure=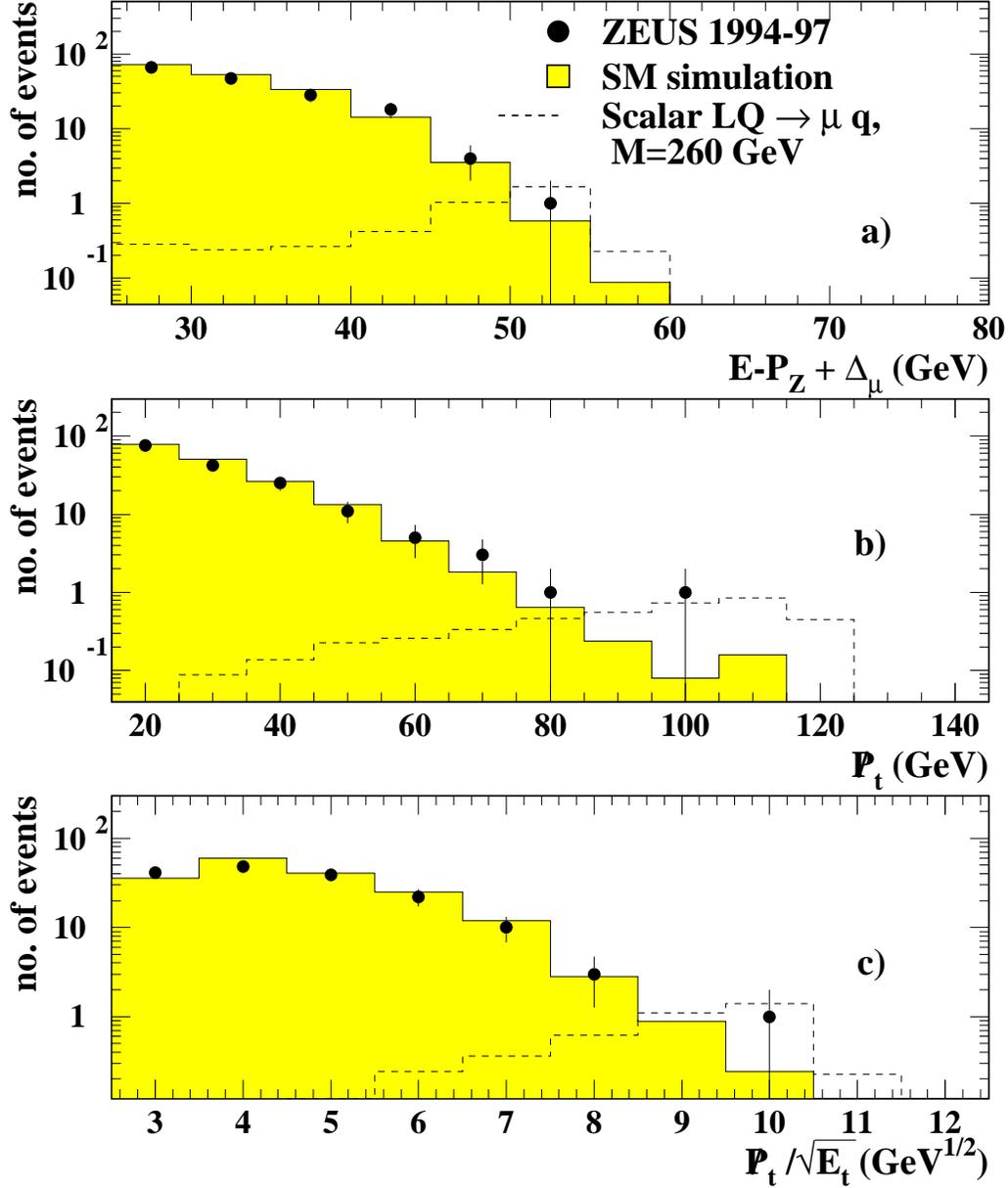,height=17.cm}
\end{center}
\caption{Distributions of event variables after the $\mu$-channel
  pre-selection for data (solid points) and SM simulation
  (shaded histograms) for a) $E-P_Z+\Delta_\mu$, b) $\ptmiss$
  and c) $\ptmiss/\sqrt{E_t}$. The
  dashed histograms simulate the signal from a scalar LQ with
  $\MLQ=260\gev$ normalized to the \CL{95}
  cross-section upper limit (see \Sect{results}).}
\label{fig-comp_cc}
\end{figure}

The third cut discriminates against photoproduction events, while the
fourth cut suppresses NC DIS. The electron finder~\cite{nim:a365:508}
is based on a neural--network algorithm.  After the pre--selection,
$164$ events remained, compared with $177.3\pm 3.8$ events
predicted by the SM simulation normalized to the integrated luminosity of the 
data. The error associated with the prediction
arises from the generated MC statistics. The SM expectation is dominated by CC DIS,
with small contributions from $ep \to e\mu^+\mu^-X$ and from $W$
production.  Simulated distributions of $E-P_Z+\Delta_\mu$, $\ptmiss$,
and $\ptmiss/\sqrt{\Et}$ are compared with the data in
\fig{comp_cc}. Good agreement is seen.

\subsection{Muon identification}
\label{sec-MuonId}
Two methods of muon identification were employed. The first, for very
forward muons ($8^\circ <\theta_\mu< 20^\circ$), required
a reconstructed track in the FMUON detector with azimuth within $20^\circ$
of $\phimiss$.
In the second selection,
for central muons ($15^\circ<\theta_\mu<164^\circ$),
the following CAL- and CTD-based requirements were imposed:
\begin{itemize}
\item a track that points to the vertex with transverse momentum ($\pttrk$)
     above 5 GeV and
     an azimuth that differs from $\phimiss$ by less than $20^\circ$;
\item no additional tracks with azimuth within $50^\circ$ of
      $\phimiss$ and $\pttrk>1\gev$;
\item the calorimeter energy deposits are consistent with
  those expected from a minimum ionizing particle
in an $(\eta,\phi)$ cone of radius $R=0.3$, centered on the track;
\item muons with $115^\circ<\theta_\mu<130^\circ$ were excluded to
eliminate the background from a very small fraction of electrons 
for which a large fraction of the energy was absorbed in the dead material
between the BCAL and the RCAL.
 \end{itemize}

After the muon identification, 2 events are left in the data, while
the SM expectation is $1.43\pm0.38$, mainly from $ep\to e\mu^+\mu^-X$.

\subsection{Final selection} 
The final selection was designed to reduce the SM background
to a very low level. The following cuts were applied:
\begin{itemize}
\item $\ptmiss> 20\gev$; 
\item $\ptmiss/\sqrt{\Et} > 4 \sqrt{\Gev}$;
\item $E-{P}_Z+\Delta_\mu>30\gev$.
\end{itemize}
No event survived these cuts, while $0.40 \pm 0.18$ events are
predicted by SM processes, mostly from $ep \to e\mu^+\mu^-X$.

\section{Event selection for the {\boldmath $e\to\tau$} transition}
\label{sec-TauEventSel}
This channel is characterized by an isolated $\tau$ with high $P_t$
balanced by a jet. Separate selections were made for hadronic $\tau$
decays ($65\%$) and for the leptonic decays
$\tau\to\ell\nubar_\ell\nu_\tau$
($35\%$). The same trigger as described in \Sect{MuEventSel} was used.
The offline event selection consisted of a pre-selection common to
hadronic and leptonic $\tau$ decays and final selections specific to each
decay mode of the $\tau$. These mode-specific selections
make use of the fact that one or more neutrinos are emitted in $\tau$ decay
producing $\ptmiss$ approximately aligned with the $\tau$.
To produce a reasonably large event sample to compare with SM predictions, the
selections for each $\tau$ decay mode were done in two steps.

\subsection{Pre-selection}\label{sec-TauPreSel}
In addition to cuts to reject non-$ep$ background,
the pre--selection requirements were:
\begin{itemize}
\item a reconstructed vertex with $Z$-coordinate $|Z_{\rm VTX}|<50$\,cm;
\item $20\gev < E-P_Z < 52\gev$;
\item energy in RCAL $<7\gev$.
\end{itemize}
The second cut reduces the photoproduction background.
The third cut rejects NC DIS events
where the positron was scattered into the RCAL.

\subsection {Selection of hadronic {\boldmath $\tau$} decays}\label{had_tau_sel}
Events with a narrow `pencil-like' jet consistent with hadronic $\tau$
decay were selected with the following requirements:
\begin{itemize}
\item the transverse energy of the jet associated with the $\tau$ should
satisfy $\Ettaujet> 10\gev$;
\item $\mjet<7\gev$;
\item 1, 2 or 3 tracks associated with the jet;
\item the number of calorimeter cells associated
  with the jet, $N_{\textrm{cells}}$, is at least 10 (to suppress electrons)
  and at most 50 (to ensure that the jet is narrow);
\item $R_{90\%}\le 0.3$,
where $R_{90\%}$ is the radius of the $(\eta,\phi)$-cone centered on
the jet axis that contains 90$\%$ of the jet energy;
\item $\femc< 0.95$,
where $\femc$ is the fraction of the jet energy
deposited in the electromagnetic section of the calorimeter;
\item $\femc+\flt< 1.6$,
where $\flt$ is the 
momentum of the most-energetic track in the jet divided by the jet energy
(leading-track fraction).
\end{itemize}

\begin{figure}[htb]
\begin{center}
\epsfig{figure=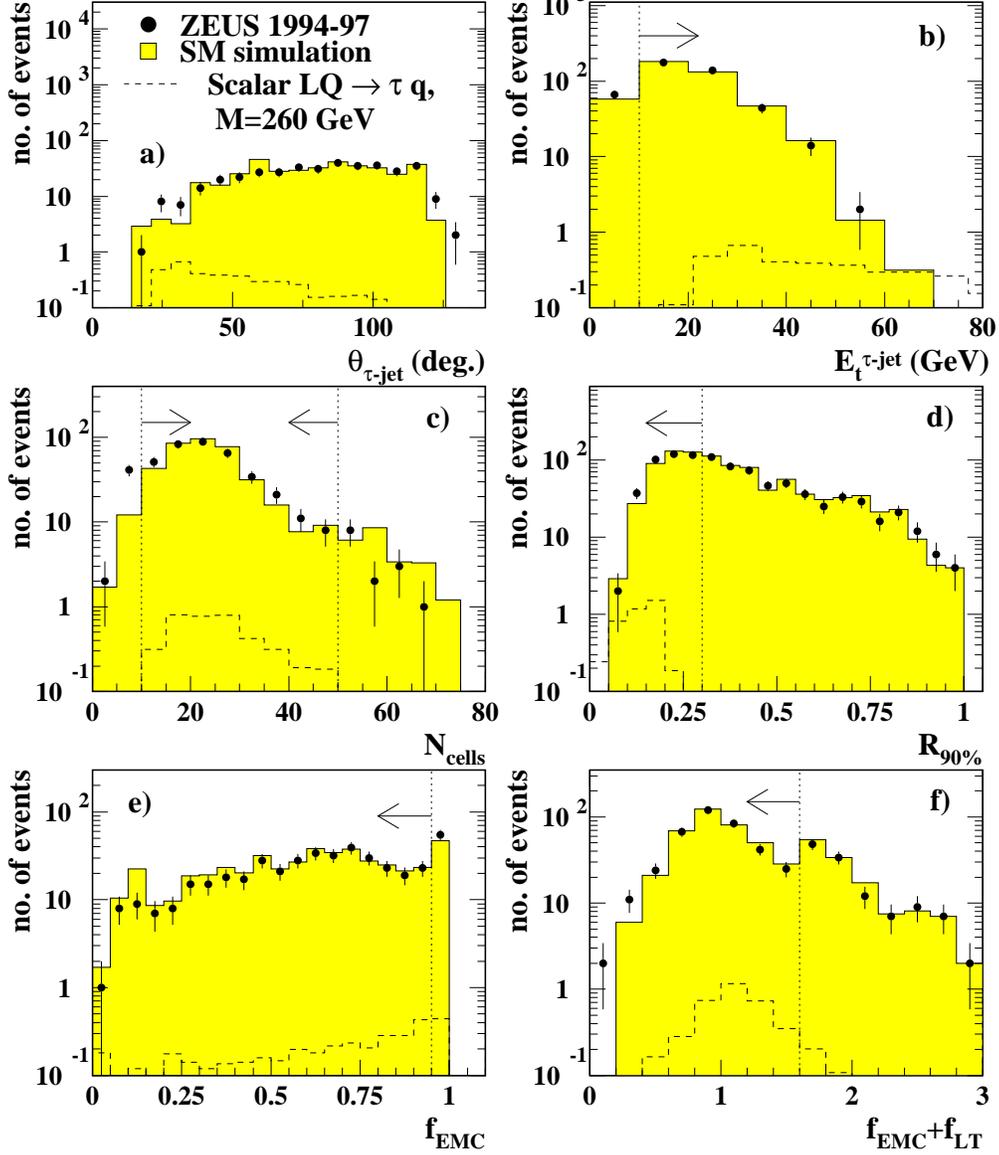,height=16.cm}
\end{center}
\caption{Comparison of data (solid points) and simulated SM background
  (shaded histograms) for candidate jets from hadronic $\tau$
  decays. The distributions are displayed for events that pass the
  selection cuts described in Section~\ref{had_tau_sel} except the
  ones imposed on the variable considered (indicated by the arrows). Shown are
the distributions of:
  (a) $\theta_{\tau\textrm{-jet}}$, the polar angle of the $\tau$ candidate jet;
  (b) $E_t^{\tau\textrm{-jet}}$, the transverse energy;
  (c) $N_{\rm{cells}}$, the number of calorimeter cells belonging to the jet;
  (d) $R_{90\%}$, the ($\eta,\phi$)-radius containing  $90\%$ of the jet energy;
  (e)$f_{\rm EMC}$, the fraction
  of the jet energy in the EMC section of the calorimeter;
  (f) $f_{\rm EMC}+f_{\rm LT}$ where
  $f_{\rm LT}$ is the momentum of the leading track divided by the jet energy.
  The SM backgrounds include  NC and CC DIS, photoproduction,
  $W$ production and $\gamma\gamma\to\tau^+\tau^-$.
  The dashed histograms simulate the signal from a scalar LQ with a mass of
  $260$${\rm{\gev}}$ normalized to the \CL{95} upper limit on the cross section
  (see \Sect{results}).}\label{fig-tau_had}  
\end{figure}

The last two cuts reject electrons, for which  $\femc\sim1$ and $\flt\sim1$.
After these cuts, 367 data events were selected in comparison to
$377.7\pm12.5$ from the SM expectation (mainly from NC DIS, CC
DIS and photoproduction). \Fig{tau_had} shows several
distributions of characteristic variables of the $\tau$ candidates at
this stage of the analysis. The SM simulation provides a reasonable
description of the data.\\ 

The final stage of the hadronic $\tau$-decay selection requires
events consistent with a two-body $\tau+$jet final state with
an invariant mass above $\sim100\gev$:
\begin{itemize}
\item $\Ettaujet> 20\gev$;
\item the azimuthal angle of the $\tau$ candidate is within $20^\circ$
of $\phimiss$;
\item $\ptmiss>12\gev$;
\item at least one additional jet with $\Etjet > 25\gev$.
\end{itemize}

No candidate satisfying these requirements was found, while
$0.62\pm0.18$ events are expected from SM processes.

\subsection{Selection of {\boldmath $\tau\to\mu{\bar\nu}_\mu\nu_\tau$} decays}
\begin{figure}[htb]
\begin{center}
\epsfig{figure=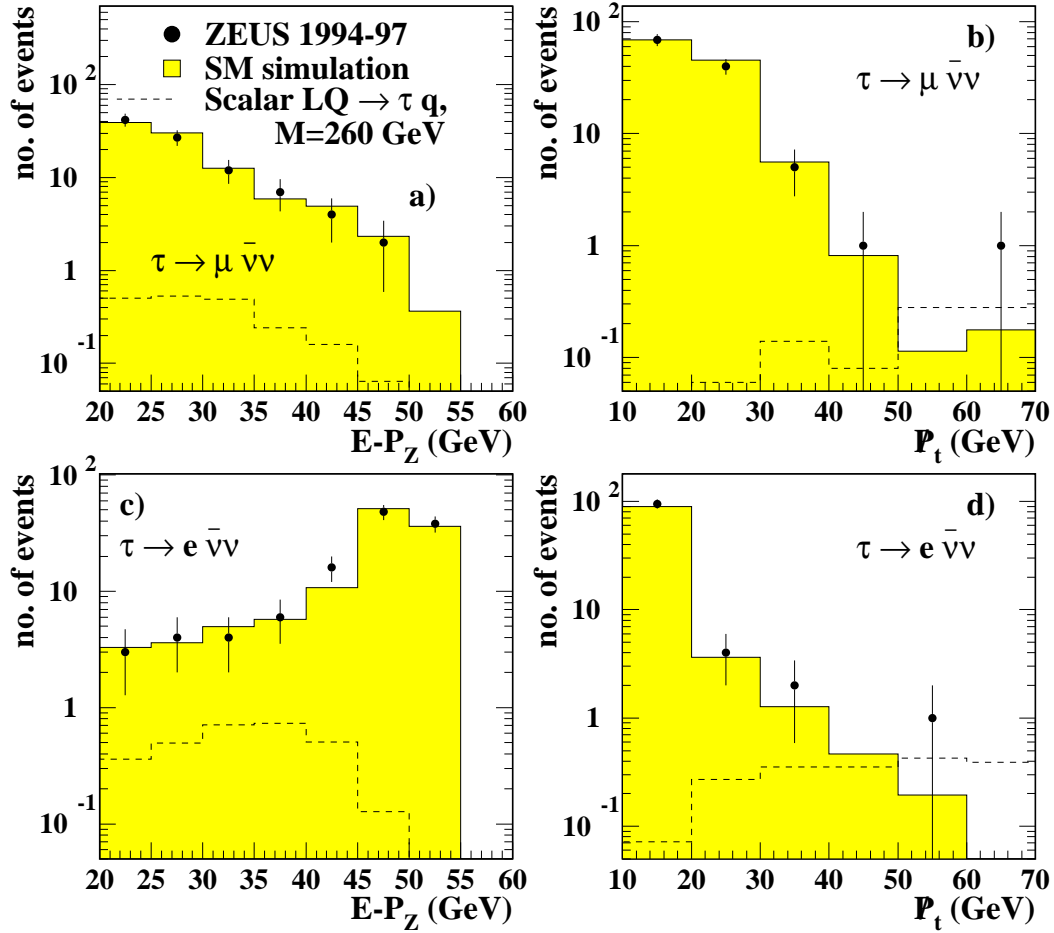,height=13.cm}
\caption{Comparison of data (solid dots) with simulated SM background
(shaded histogram) for the distributions of a) $E-P_Z$ and b)
$\ptmiss$ for the $\tau \to \mu \nubar \nu$ selection.
The same distributions for the $\tau \to e \nubar \nu$ selection
are shown in c) and d), respectively.
The SM backgrounds include NC DIS, photoproduction, CC DIS, $W$ production
and $\gamma\gamma\to\tau^+\tau^-$.
The dashed histograms simulate the signal from a scalar LQ with a mass of
 $260\gev$ normalized to the \CL{95} upper limit on the cross section
 (see \Sect{results}).}\label{fig-tau_lep} 
\end{center}
\end{figure}

After the pre-selection described in \Sect{TauPreSel}, events with an
isolated high-$P_t$ muon balanced by a jet were selected.  Isolated
muon candidates were identified using a neural-network algorithm that
analyzed the pattern of longitudinal and transverse energy deposition
in the calorimeter and matching track(s) in the CTD and/or the muon
chambers.  Since the energy deposited in the CAL by the muon is
typically a small fraction of the energy of the $\tau$, cuts on
$\ptmiss$ were applied. The initial requirements were:
\begin{itemize}
\item a muon with $P_t>10\gev$ and $8^\circ<\theta_\mu<125^\circ$;
\item $\ptmiss>15\gev$;
\item $\ptmiss/{\sqrt{\Et}}>4\sqrt{\Gev}$;
\item a jet with $\Etjet>25\gev$;
\item events with an identified electron~\cite{epj:c11:427} with energy
greater than $10\gev$ were vetoed.
\end{itemize}
The last cut reduces NC DIS background.
\Figs{tau_lep}(a-b) show the distributions of $E-P_Z$ and
$\ptmiss$ for the $\tau\to\mu\nubar\nu$ candidates after these cuts,
compared to the SM background. Good agreement is observed. After these 
cuts, 119 data events remained,
compared to $107.2\pm7.4$ events from the SM expectation
(mainly CC DIS and photoproduction). 

The final selection consisted of two cuts:
\begin{itemize}
\item $\ptmiss>20\gev$;
\item the muon azimuth differs from $\phimiss$ by less than  $20^\circ$.
\end{itemize}

No event passed the final selection, while $0.23\pm0.07$
events were expected from SM processes.

\subsection{Selection of {\boldmath $\tau\to e{\bar\nu}_e\nu_\tau$} decays}
After the pre-selection described in \Sect{TauPreSel},
events with an isolated electron and a jet were selected by imposing
the following requirements:
\begin{itemize} 
\item an identified electron~\cite{epj:c11:427} with energy greater than
$20\gev$ and the polar angle $\theta_e$ satisfying $8^\circ<\theta_e<125^\circ$;
\item $\ptmiss>10\gev$;
\item $\ptmiss/\sqrt{\Et}>2\sqrt{\Gev}$;
\item a jet with $\Etjet>25\gev$.
\end{itemize}

\Figs{tau_lep}(c-d) show the
distributions of $E-P_Z$ and $\ptmiss$ for the $\tau\to e\nubar\nu$
candidates that satisfied these requirements, where 116 data events
were selected and $109.1\pm 5.4$ events from SM backgrounds were expected
(mainly NC DIS and photoproduction).

The final selection consisted of a higher $\ptmiss$ cut and a requirement
that the lepton and the jet be back-to-back:
\begin{itemize}
\item $\ptmiss>15\gev$;
\item the azimuth of the electron differs from $\phimiss$ by less than  $20^\circ$.
\end{itemize}

No event passed the final selection, while $0.32\pm0.10$
events were expected from SM processes.

\section{Efficiencies}

The selection efficiencies were evaluated using signal MC events
(see \Sect{MC}). For resonant production of lepton-flavor-violating
scalar LQs, the $\mu$-channel selection efficiency 
falls from 60$\%$ to 52$\%$ as $\MLQ$ increases from 140$\gev$ to 280$\gev$,
while the efficiency for vector LQs drops from
64$\%$ to 56$\%$. 
For $\MLQ>240\gev$, the FMUON-based muon selection increases the selection
efficiency by about $20\%$ compared to the CAL-CTD-based selection alone.
Over the $\MLQ$ interval from 140$\gev$ to 280$\gev$,
 the selection efficiency for LQs that couple to $\tau$ increases
from 24$\%$ to 31$\%$ for scalar LQs and from 21$\%$ to 33$\%$ for
vector LQs.

For LQs with $\MLQ\gg\sqrt{s}$, the efficiencies are almost independent 
of $\MLQ$,
but depend strongly on the generation of the initial-state quark.
For $e\to \mu$ transitions, the selection efficiency ranges from
$15\%$ to $45\%$ for $F=0$ LQs and from $15\%$ to $35\%$
for $|F|=2$ LQs. For $e\to \tau$ transitions, the efficiencies are lower
and range from $5\%$ to $19\%$ for $F=0$ LQs and from $4\%$ to $16\%$
for $|F|=2$ LQs. When the initial-state quark is a sea quark and
especially for $s$, $c$, or $b$ quarks,
the efficiency is considerably lower than for valence quarks 
due to the softer $x$ spectrum,
which results in a lower transverse momentum for the final-state lepton.


\section{Results}\label{sec-results}
Since no candidate for LFV processes was found, limits
were set on these processes. 
All limits were evaluated at \CL{95} using a Bayesian approach, assuming
a flat prior for the signal cross section. Systematic
uncertainties in the detector simulation and in the 
integrated luminosity (see \Sect{syst})
were taken into account using a method described
elsewhere~\cite{proc:clw:2000:237}.
For low-mass LQs with narrow width, the branching ratio, $\br{}$, was
regarded as a free parameter and limits were set on $\sigma\br{\ell\qflow}$.
These limits were converted to limits on 
$\lambda_{e \qone}\sqrt{\br{\ell\qflow}}$
using \eq{lowmas} corrected for QED-ISR and NLO QCD (only for scalar LQs).
For high-mass leptoquarks, the cross-section limit was converted
to a limit on  $\lambda_{e \qe}\lambda_{\ell\ql}/\MLQ^2$ using
\eq{highmas} with QED-ISR corrections.
The CTEQ4~\cite{pr:d55:1280} parameterizations of parton densities
were used to evaluate cross sections.

\subsection{Systematic uncertainties}\label{sec-syst}

The uncertainty on the integrated luminosity is $1.6\%$.
Systematic uncertainties of $3\%$ on the CAL energy scale
and $10\%$ on the CAL response to muons were taken into account.
The resulting variations on the efficiency for
the muon (tau) channel were $3\%$ ($4\%$) for low-mass LQs and up to
$15\%$ ($17\%$) for high-mass LQs that couple to $b$ quarks in the initial state. 

Systematic uncertainties in the cross-section evaluation, related 
to the choice of parton density function (PDF), were investigated using 
MRST~\cite{epj:c4:463-tmp-3be95bca} as an alternative choice to CTEQ4.
The main differences were found for low-mass LQs with masses close to 
$\sqrt{s}$ when
very high-$x$ quarks are involved. In these cases, limits calculated using 
MRST were stricter than the CTEQ4-based limits presented here. 
Another possible source of uncertainties for vector 
and high-mass scalar LQs are the unknown NLO-QCD cross-section corrections  
(see \Sect{theory}).

\subsection{Low-mass LQ and squark limits}
\label{sec-lowlim}
\begin{figure}[htb]
\begin{center}
\epsfig{figure=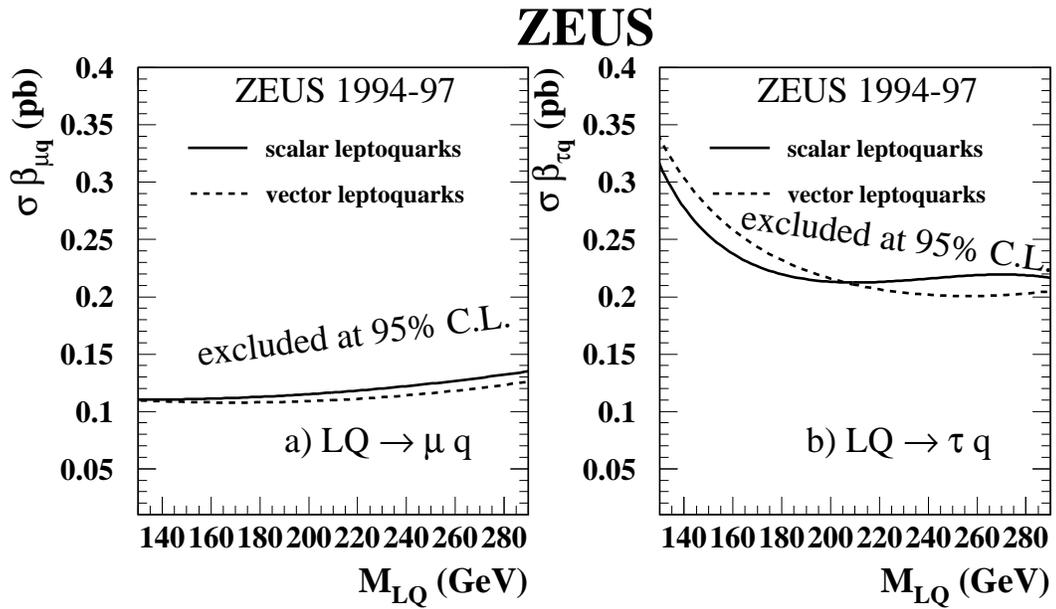,height=8.cm}
\end{center}
\caption{The \CL{95} upper limits on $\sigma\br{\ell q}$ as a
  function of $\MLQ$ for scalar (full line) and vector (dashed
  line) LQs for a) $\LQ\to\mu q$ and b) $\LQ\to\tau q$.}\label{fig-sigma_br}
\end{figure}

\Fig{sigma_br} shows upper limits on $\sigma\br{\mu\qflow}$
and $\sigma\br{\tau\qflow}$.  For $e\to\mu$, the search is sensitive to
processes with cross sections as low as $0.1\pb$, while for $e\to\tau$,
the sensitivity is $0.2\rnge0.3\pb$. These limits apply generally to narrow
resonances with LFV decay modes, for example, to the \rp\, squarks
described in \Sect{susy}.

Upper limits on  $\lambda_{e\qone}\sqrt{\br{\ell\qflow}}$
have been derived for F=0 LQs
by assuming resonantly produced LQs described by the BRW model.
These limits can be applied to
processes involving any quark generations in the final state (excluding the
$t-$quark). 
\Figs{mu_low}(a-b) show the upper limits on
$\lambda_{e\qone} \sqrt{\br{\mu\qflow}}$
for scalar and vector LQs, respectively.
Under the assumption that $\lambda_{e\qone}=\lambda_{\mu\qflow_\beta}$,
limits on $\lambda_{e\qone}$ can be derived. These are compared
to limits from low-energy experiments in
\figs{mu_low}(c-d) for $\tShL$ and $\VzR$ LQs.
These states do not couple to neutrinos and therefore $\br{\mu q}=0.5$.
For $\MLQ<250\gev$, the ZEUS limits are stronger than the low-energy
limits for LQs that couple to $\mu$ and $b$.
Limits on $\lambda_{e\qone}\sqrt{\br{\tau\qflow}}$ are shown
in \figs{tau_low}(a-b). \Figs{tau_low}(c-d)
show the corresponding limits on $\lambda_{e\qone}$, 
assuming $\lambda_{e \qone} = \lambda_{\tau\qflow_\beta}$
(and therefore $\br{\tau q}=0.5$).
The ZEUS limits are more stringent
than the limits from low-energy experiments over a wide mass range,
with the exception of limits 
from $K^+ \to \pi^+ \nu \nubar$~\cite{prl:84:3768}.
As described in \Sect{susy}, the limits
on $\lambda_{e \qone}\sqrt{\br{\ell\qflow}}$ for
$\tilde{S}_{1/2}^L$  can be interpreted as limits on
$\lambda_{1j1}^\prime\sqrt{\br{\tilde u^j \to \ell\qflow}}$ for
$\tilde{u}^j$ squarks.

Another way to illustrate the sensitivity is to assume that
the couplings have electromagnetic
strength ($\lambda_{e\qone}=\lambda_{\ell\qflow_\beta}=0.3\approx\sqrt{4\pi\alpha}$).
In this case, LQs with masses up to $283\gev$ are
excluded, as shown in \Tab{low_mas_limit}. 
Alternatively, as shown in \Tab{low_coupl_limit},
for a fixed $\MLQ$ of 250 GeV, values of
$\lambda_{e\qone}\sqrt{\br{\ell\qflow}}$ down
to $0.020$ ($0.027$) for $LQ\to\mu q$ ($LQ\to \tau q$)
are excluded.

The CDF~\cite{prl:81:4806} and D{\O}~\cite{prl:84:2088} 
collaborations exclude
scalar LQs with $\MLQ<202\gev$ and
$\MLQ<200\gev$, respectively, at \CL{95} with $\br{\mu\qflow}=100\%$.
CDF~\cite{prl:78:2906} excludes $\MLQ<99\gev$ with $\br{\tau b}=100\%$.
The ZEUS limits are complementary to those of the Tevatron in
the sense that the latter are independent of the Yukawa couplings and
assume that LQs couple to a single lepton generation.

\subsection{High-mass LQ and squark limits}\label{sec-himassLQ}
\label{sec-higlim}
For $\MLQ\gg\sqrt{s}$, limits on $\lambda_{e\qe} \lambda_{\ell  \ql}/\MLQ^2$
were evaluated for all combinations of quark
generations ($\alpha$, $\beta$). \taband{HMF0}{HMF2} 
show these limits for $F=0$ and $|F|=2$ LQs, respectively, that couple to
$\mu \ql$.
\taband{F0}{F2} show
the corresponding limits for the LQs coupling to $\tau \ql$.
In many cases involving $c$ and $b$ quarks, the ZEUS limits improve
on the low-energy limits~\cite{zfp:c61:613,pr:d62:055009,epj:c15:1}.
Limits obtained by H1~\cite{epj:c11:447} are comparable
to the ZEUS limits.

The limits on $\lambda_{e\qe} \lambda_{\ell  \ql}/\MLQ^2$ for
 $\tilde{S}_{1/2}^L$ can be interpreted as limits
on $\lambda'_{1j\alpha}\lambda'_{ij\beta}/M_{\tilde u}^2$ for a $u$-type
squark of generation $j$, where $\ell=\mu$ or $\tau$ for $i=2$ or 3, respectively.
Similarly, the limits on $S_0^L$ LQs can be interpreted as limits
on $\lambda'_{1\alpha k}\lambda'_{i\beta k}/M_{\tilde d}^2$ for a $d$-type squark of
generation $k$.

\section{Conclusions}
A search for lepton-flavor violation has been performed with $47.7\pbi$ of
$e^+p$ data at $\sqrt{s}=300 \gev$ collected with the ZEUS
detector at HERA in 1994--1997. Both the $\mu$ and $\tau$ channels have
been analyzed. No evidence for LFV processes has been found. 

Limits at \CL{95} on
cross sections, couplings and masses for F=0 LQs that mediate
LFV processes have been set. Assuming the couplings
$\lambda_{e\qone}=\lambda_{\ell\qflow_\beta}=0.3$,
lower mass limits between 258 and 283 GeV
have been derived for various LQs decaying to $\mu q$ or $\tau q$.
For $\MLQ=250\gev$, upper limits for
$\lambda_{e\qone}\sqrt{\br{\mu\qflow}}$ in the range
$(2.0\rnge 10)\cdot10^{-2}$ and
for $\lambda_{e\qone}\sqrt{\br{\tau\qflow}}$ in the range
$(2.7\rnge 15)\cdot10^{-2}$ were obtained.
Limits on ${\tilde S}^L_{1/2}$ also apply to
up-type squarks that have $R$-parity-violating couplings
to both a positron and either a $\mu$ or a $\tau$.

For LQs with $\MLQ \gg \sqrt{s}$, upper limits on 
$\lambda_{e \qe}\lambda_{\ell  \ql}/{\MLQ^2}$
have been obtained and compared with
bounds from low-energy experiments.
Some of these limits also apply to high-mass
$R_p$-violating squarks.
A number of ZEUS limits
are the most stringent published to date, especially for $e\to\tau$
transitions.

\section*{Acknowledgements}
    
The strong support and encouragement of the DESY directorate have been
invaluable. The experiment was made possible by the inventiveness and
the diligent efforts of the HERA machine group. The design,
construction and installation of the ZEUS detector have been made
possible by the ingenuity and dedicated efforts of many people from
inside DESY and from the home institutes who are not listed as
authors. Their contributions are acknowledged with great appreciation.
We acknowledge support by the following:
the Natural Sciences and Engineering Research Council of Canada (NSERC);
the German Federal Ministry for Education and
Science, Research and Technology (BMBF), under contract
numbers HZ1GUA 2, HZ1GUB 0, HZ1PDA 5, HZ1VFA 5;
the MINERVA Gesellschaft f\"ur Forschung GmbH, the
Israel Science Foundation, the U.S.-Israel Binational Science
Foundation, the Israel Ministry of Science and the Benozyio Center
for High Energy Physics;
the German-Israeli Foundation, the Israel Science
Foundation, and the Israel Ministry of Science;
the Italian National Institute for Nuclear Physics (INFN);
the Japanese Ministry of Education, Science and
Culture (the Monbusho) and its grants for Scientific Research;
the Korean Ministry of Education and Korea Science
and Engineering Foundation;
the Netherlands Foundation for Research on Matter (FOM);
the Polish State Committee for Scientific Research,
grant no. 115/E-343/SPUB-M/DESY/P-03/DZ 121/2001-2002
and by the German Federal Ministry for
Education and Science, Research and Technology (BMBF);
the Fund for Fundamental Research of Russian
Ministry for Science and Edu\-cation and by the German Federal Ministry for
Education and Science, Research and Technology (BMBF);
the Spanish Ministry of Education
and Science through funds provided by CICYT;
the Particle Physics and Astronomy Research Council, UK;
the US Department of Energy;
the US National Science Foundation.


{
\def\bibname{\Large\bf References}
\def\refname{\Large\bf References}
\pagestyle{plain}
\ifzeusbst
  \bibliographystyle{./BiBTeX/bst/l4z_default}
\fi
\ifzdrftbst
  \bibliographystyle{./BiBTeX/bst/l4z_draft}
\fi
\ifzbstepj
  \bibliographystyle{./BiBTeX/bst/l4z_epj}
\fi
\ifzbstnp
  \bibliographystyle{./BiBTeX/bst/l4z_np}
\fi
\ifzbstpl
  \bibliographystyle{./BiBTeX/bst/l4z_pl}
\fi
{\raggedright
\bibliography{./BiBTeX/user/syn.bib,%
              ./BiBTeX/bib/l4z_articles.bib,%
              ./BiBTeX/bib/l4z_books.bib,%
              ./BiBTeX/bib/l4z_conferences.bib,%
              ./BiBTeX/bib/l4z_h1.bib,%
              ./BiBTeX/bib/l4z_misc.bib,%
              ./BiBTeX/bib/l4z_old.bib,%
              ./BiBTeX/bib/l4z_preprints.bib,%
              ./BiBTeX/bib/l4z_replaced.bib,%
              ./BiBTeX/bib/l4z_temporary.bib,%
              ./BiBTeX/bib/l4z_zeus.bib}}
}

\newcommand{\himasscaptionA}[2]{\caption{Upper limits at \CL{95}
  on $\lambda_{e\qe}\lambda_{#1\ql}/M^2_{LQ}$ in units of
  ${\rm TeV}^{-2}$, for $#2$ LQs that couple to $e\qe$ and to $#1\ql$.
  The columns correspond to the $#2$ LQ species. The $eq_\alpha$
  combination for the $s$-channel case is reported under the LQ type.
  Each row corresponds to
  a different combination of quark generations $(\alpha, \beta)$ which couple
  to the positron and the $#1$, respectively.
  Within each cell, the measurement which provides the most stringent 
  low--energy constraint is shown on the first line and the corresponding
  limit~\pcite{zfp:c61:613,pr:d62:055009,epj:c15:1} is given on the second line.
  The ZEUS limits are shown on the third line of each cell (enclosed in a
  box when stronger than the low--energy constraint).
  The * indicates cases where a top quark must be involved.}}
\newcommand{\himasscaptionB}[2]{\caption{Upper limits at \CL{95}
  on $\lambda_{e\qe}\lambda_{#1\ql}/M^2_{LQ}$ in units of
  ${\rm TeV}^{-2}$, for $#2$ LQs that couple to $e\qe$ and to $#1\ql$.
  The columns correspond to the $#2$ LQ species. 
  The format of the table is described in the caption of \Tab{HMF0}.}}
\newcommand{\Zbox}[1]{\framebox{\bf #1}}
\newcommand{\sci}[2]{$#1\times10^{#2}$}
\newcommand{\bm}[1]{{\boldmath $#1$}}
\newcommand{\lbm}[1]{\large{\boldmath $#1$}}
\newcommand{\Zn}[1]{{\boldmath $#1$}}
\newcommand{\eu}{$e^+u$}
\newcommand{\eua}{$e^+u_\alpha$}
\newcommand{\eub}{$e^+\bar{u}$}
\newcommand{\euba}{$e^+\bar{u}_\alpha$}
\newcommand{\ed}{$e^+d$}
\newcommand{\eda}{$e^+d_\alpha$}
\newcommand{\edb}{$e^+\bar{d}$}
\newcommand{\edba}{$e^+\bar{d}_\alpha$}
\newcommand{\eud}{$e^+(u+d)$}
\newcommand{\euda}{$e^+(u+d)_\alpha$}
\newcommand{\eudb}{$e^+(\bar{u}+\bar{d})$}
\newcommand{\eudba}{$e^+(\bar{u}+\bar{d})_\alpha$}
\newcommand{\euud}{$e^+ (\sqrt{2} u + d) $}
\newcommand{\euuda}{$e^+ (\sqrt{2} u + d)_\alpha$}
\newcommand{\euddb}{$e^+ (\bar{u}+\sqrt{2}\bar{d})$}
\newcommand{\euddba}{$e^+ (\bar{u}+\sqrt{2}\bar{d})_\alpha$}
\newcommand{\muNeN}{\bm{\mu N \rightarrow e N}}
\newcommand{\Bmue}{\bm{B \rightarrow \mu \bar{e}}}
\newcommand{\BmueK}{\bm{B \rightarrow \bar{\mu} e K}}
\newcommand{\Dmue}{\bm{D \rightarrow \mu \bar{e}}}
\newcommand{\Kmue}{\bm{K \rightarrow \mu \bar{e}}}
\newcommand{\mueee}{\bm{\mu \to ee\bar{e}}}
\newcommand{\Kpinunu}{\bm{K \rightarrow \pi \nu \bar{\nu}}}
\newcommand{\BlnuX}{\bm{B \rightarrow l \nu X}}
\newcommand{\taupie}{\bm{\tau \to \pi e}}
\newcommand{\tauKe}{\bm{\tau \to K e}}
\newcommand{\BtaueX}{\bm{B\to\tau\bar{e} X}}
\newcommand{\taueee}{\bm{\tau \to ee\bar{e}}}
\newcommand{\alphabeta}{\lbm{\alpha \beta}}


\vspace*{3cm}
\begin{table}[htb]
\begin{center}
\begin{tabular}{| c || c | c | c | c | c | c | c | c |} \hline
LQ type & $\tilde S_{1/2}^L$ & $S_{1/2}^L$ & $S_{1/2}^R$ & $V_0^L$ & $V_0^R$ & $\tilde V_0^R$ & $V_1^L$ \\ \hline \hline
$\mu$-channel limit on $\MLQ [\gev]$ & 263 & 278 & 278 & 261 & 266 & 280 & 283 \\ \hline
$\tau$-channel limit on $\MLQ [\gev]$ & 258 & 275 & 276 & 259 & 263 & 277 & 282 \\ \hline
\end{tabular}
\caption{ The \CL{95} lower limits on $\MLQ$ for the $\mu$- and the
$\tau$-channels assuming $\lambda_{e\qone}=\lambda_{\ell\qflow}=0.3$.}
\label{tab-low_mas_limit} 
\end{center}
\end{table}
\vspace*{5cm}
\begin{table}[htb]
\begin{center}
\begin{tabular}{| c || c | c | c | c | c | c | c |} \hline
LQ type & $\tilde S_{1/2}^L$ & $S_{1/2}^L$ & $S_{1/2}^R$ & $V_0^L$/$V_0^R$ & $\tilde V_0^R$ & $V_1^L$ \\ \hline \hline
$\mu$-channel limit on $\lambda_{e \qone}\sqrt{\br{\mu\qflow}}$  & 0.10 & 0.038 & 0.036 & 0.081 & 0.029 & 0.020 \\ \hline
$\tau$-channel limit on $\lambda_{e \qone}\sqrt{\br{\tau\qflow}}$& 0.15 & 0.054 & 0.051 & 0.10  & 0.038 & 0.027 \\ \hline
\end{tabular}
\caption{\it The \CL{95} upper limits on
  $\lambda_{e\qone}\sqrt{\br{\ell\qflow}}$ for a
  leptoquark with mass $\MLQ=250{\rm \gev}$.}\label{tab-low_coupl_limit}
\end{center}
\end{table}
\vspace{-40cm}
\begin{center}
\begin{table}[htb]
\footnotesize{
\begin{tabular}{|c||c|c|c|c|c|c|c|} \hline
\multicolumn{4}{|c}{} & \multicolumn{4}{c|}{} \\ 
 \multicolumn{1}{|c}{} &\multicolumn{2}{c}{\large{$e \to \mu$}} &\multicolumn{2}
{c}{\Large{ZEUS}} & \multicolumn{2}{c}{\large{$F=0$}} & \\ 
\multicolumn{4}{|c}{} & \multicolumn{4}{c|}{} \\ \hline
                 &               &               &               &               &               &               &               \\
\alphabeta       & \lbm{\ShL}    & \lbm{\ShR}    & \lbm{\tShL}   & \lbm{\VzL}    & \lbm{\VzR}    & \lbm{\tVzR}   & \lbm{\VoL}    \\ 
                 & \eua           & \euda          & \eda           & \eda           & \eda           & \eua           & \euuda         \\ \hline \hline
                 & \muNeN        & \muNeN        & \muNeN        & \muNeN        & \muNeN        & \muNeN        & \muNeN        \\ 
  1 1            & \sci{7.6}{-5} & \sci{2.6}{-5} & \sci{7.6}{-5} & \sci{2.6}{-5} & \sci{2.6}{-5} & \sci{2.6}{-5} & \sci{1.1}{-5} \\     
                 & \Zn{1.9}      & \Zn{1.6}      & \Zn{2.9}      & \Zn{1.9}      & \Zn{1.9}      & \Zn{1.5}      & \Zn{0.7}      \\ \hline
                 & \Dmue         & \Kmue         & \Kmue         & \Kmue         & \Kmue         & \Dmue         & \Kmue         \\ 
  1 2            & $4$           & \sci{2.7}{-5} & \sci{2.7}{-5} & \sci{1.3}{-5} & \sci{1.3}{-5} & $2$           & \sci{1.3}{-5} \\     
                 & \Zbox{1.9}    & \Zn{1.6}      & \Zn{3.0}      & \Zn{2.3}      & \Zn{2.3}      & \Zbox{1.7}    & \Zn{0.8}      \\ \hline
                 &               & \Bmue         & \Bmue         & $\mathbf V_{ub}$      & \Bmue         &               & $\mathbf V_{ub}$      \\ 
  1 3            & \Zn{*}        & $0.8$         & $0.8$         & $0.2$         & $0.4$         & \Zn{*}        & $0.2$         \\     
                 &               & \Zn{3.1}      & \Zn{3.1}      & \Zn{2.7}      & \Zn{2.7}      &               & \Zn{2.7}      \\ \hline
                 & \Dmue         & \Kmue         & \Kmue         & \Kmue         & \Kmue         & \Dmue         & \Kmue         \\ 
  2 1            & $4$           & \sci{2.7}{-5} & \sci{2.7}{-5} & \sci{1.3}{-5} & \sci{1.3}{-5} & $2$           & \sci{1.3}{-5} \\
                 & \Zn{8.5}      & \Zn{4.9}      & \Zn{6.2}      & \Zn{2.8}      & \Zn{2.8}      & \Zn{3.2}      & \Zn{1.5}      \\ \hline
                 & \mueee        & \mueee        & \mueee        & \mueee        & \mueee        & \mueee        & \mueee        \\ 
  2 2            & \sci{5}{-3}   & \sci{7.3}{-3} & \sci{1.6}{-2} & \sci{8}{-3}   & \sci{8}{-3}   & \sci{2.5}{-3} & \sci{1.5}{-3} \\
                 & \Zn{11}       & \Zn{5.5}      & \Zn{6.9}      & \Zn{3.4}      & \Zn{3.4}      & \Zn{5.1}      & \Zn{2.2}      \\ \hline
                 &               & \BmueK        & \BmueK        & \BmueK        & \BmueK        &               & \BmueK        \\ 
  2 3            & \Zn{*}        & $0.6$         & $0.6$         & $0.3$         & $0.3$         & \Zn{*}        & $0.3$         \\     
                 &               & \Zn{8.8}      & \Zn{8.8}      & \Zn{5.7}      & \Zn{5.7}      &               & \Zn{5.7}      \\ \hline
                 &               & \Bmue         & \Bmue         & $\mathbf V_{ub}$      & \Bmue         &               & $\mathbf V_{ub}$      \\ 
  3 1            &  \Zn{*}       & $0.8$         & $0.8$         & $0.2$         & $0.4$         & \Zn{*}        & $0.2$         \\     
                 &               & \Zn{9.3}      & \Zn{9.3}      & \Zn{3.2}      & \Zn{3.2}      &               & \Zn{3.2}      \\ \hline
                 &               & \BmueK        & \BmueK        & \BmueK        & \BmueK        &               & \BmueK        \\ 
  3 2            & \Zn{*}        & $0.6$         & $0.6$         & $0.3$         & $0.3$         & \Zn{*}        & $0.3$         \\     
                 &               & \Zn{11}       & \Zn{11}       & \Zn{3.9}      & \Zn{3.9}      &               & \Zn{3.9}      \\ \hline
                 &               & \mueee        & \mueee        & \mueee        & \mueee        &               & \mueee        \\            
  3 3            & \Zn{*}        & \sci{7.3}{-3} & \sci{1.6}{-2} & \sci{8}{-3}   & \sci{8}{-3}   & \Zn{*}        & \sci{1.5}{-3} \\
                 &               & \Zn{16}       & \Zn{16}       & \Zn{8.0}      & \Zn{8.0}      &               & \Zn{8.0}      \\ \hline 
\end{tabular}}
\himasscaptionA{\mu}{F=0}
\label{tab-HMF0}
\end{table}
\end{center}

\begin{center}
\begin{table}[htb]
\footnotesize{
\begin{tabular}{|c||c|c|c|c|c|c|c|} \hline
\multicolumn{4}{|c}{} & \multicolumn{4}{c|}{} \\ 
\multicolumn{1}{|c}{}&\multicolumn{2}{c}{\large{$e \to \mu$}} &\multicolumn{2}
{c}{\Large{ZEUS}} & \multicolumn{2}{c}{\large{$|F|=2$}} &\\ 
\multicolumn{4}{|c}{} & \multicolumn{4}{c|}{} \\ \hline
                 &               &               &               &               &               &               &               \\
\alphabeta       & \lbm{\SzL}    & \lbm{\SzR}    & \lbm{\tSzR}   & \lbm{\SoL}    & \lbm{\VhL}    & \lbm{\VhR}    & \lbm{\tVhL}   \\ 
                 & \euba          & \euba          & \edba          & \euddba        & \edba          & \eudba         & \euba          \\ \hline \hline
                 & \muNeN        & \muNeN        & \muNeN        & \muNeN        & \muNeN        & \muNeN        & \muNeN        \\ 
    1 1          & \sci{7.6}{-5} & \sci{7.6}{-5} & \sci{7.6}{-5} & \sci{2.3}{-5} & \sci{2.6}{-5} & \sci{1.3}{-5} & \sci{2.6}{-5} \\     
                 & \Zn{3.4}      & \Zn{3.4}      & \Zn{4.2}      & \Zn{1.8}      & \Zn{1.5}      & \Zn{0.8}      & \Zn{1.0}      \\ \hline
                 & \Kpinunu      & \Dmue         & \Kmue         & \Kmue         & \Kmue         & \Kmue         & \Dmue         \\ 
    1 2          & $10^{-3}$     & $4$           & \sci{2.7}{-5} & \sci{1.3}{-5} & \sci{1.3}{-5} & \sci{1.3}{-5} & $2$           \\     
                 & \Zn{7.1}      & \Zn{7.1}      & \Zn{5.6}      & \Zn{2.6}      & \Zn{3.1}      & \Zn{2.5}      & \Zn{4.4}      \\ \hline
                 & $\mathbf V_{ub}$      &               & \Bmue         & $\mathbf V_{ub}$      & \Bmue         & \Bmue         &               \\ 
    1 3          & $0.4$         &  \Zn{*}       & $0.8$         & $0.4$         & $0.4$         & $0.4$         & \Zn{*}        \\     
                 & \Zn{*}        &               & \Zn{6.6}      & \Zn{3.2}      & \Zn{4.7}      & \Zn{4.7}      &               \\ \hline
                 & \Kpinunu      & \Dmue         & \Kmue         & \Kmue         & \Kmue         & \Kmue         & \Dmue         \\ 
    2 1          & $10^{-3}$     & $4$           & \sci{2.7}{-5} & \sci{1.3}{-5} & \sci{1.3}{-5} & \sci{1.3}{-5} & $2$           \\     
                 & \Zn{3.7}      & \Zbox{3.7}    & \Zn{4.7}      & \Zn{2.0}      & \Zn{1.6}      & \Zn{0.9}      & \Zbox{1.0}    \\ \hline
                 & \mueee        & \mueee        & \mueee        & \mueee        & \mueee        &  \mueee       & \mueee        \\
    2 2          & \sci{5}{-3}   & \sci{5}{-3}   & \sci{1.6}{-2} & \sci{1.3}{-2} & \sci{8}{-3}   & \sci{3.7}{-3} & \sci{2.5}{-3} \\
                 & \Zn{11}       & \Zn{11}       & \Zn{6.9}      & \Zn{3.4}      & \Zn{3.4}      & \Zn{2.8}      & \Zn{5.1}      \\ \hline
                 & \BlnuX        &               & \BmueK        & \BmueK        & \BmueK        & \BmueK        &               \\ 
    2 3          & $4$           & \Zn{*}        & $0.6$         & $0.3$         & $0.3$         & $0.3$         & \Zn{*}        \\     
                 & \Zn{*}        &               & \Zn{8.8}      & \Zn{4.4}      & \Zn{5.7}      & \Zn{5.7}      &               \\ \hline
                 & $\mathbf V_{ub}$      &               & \Bmue         & $\mathbf V_{ub}$      & \Bmue         & \Bmue         &               \\ 
    3 1          & $0.4$         & \Zn{*}        & $0.8$         & $0.4$         & $0.4$         & $0.4$         & \Zn{*}        \\     
                 & \Zn{*}        &               & \Zn{5.6}      & \Zn{2.8}      & \Zn{1.6}      & \Zn{1.6}      &               \\ \hline
                 & \BlnuX        &               & \BmueK        & \BmueK        & \BmueK        & \BmueK        &               \\ 
    3 2          & $4$           & \Zn{*}        & $0.6$         & $0.3$         & $0.3$         & $0.3$         & \Zn{*}        \\     
                 & \Zn{*}        &               & \Zn{11}       & \Zn{5.6}      & \Zn{3.9}      & \Zn{3.9}      &               \\ \hline
                 &               &               & \mueee        & \mueee        & \mueee        & \mueee        &               \\ 
    3 3          & \Zn{*}        & \Zn{*}        & \sci{1.6}{-2} & \sci{1.3}{-2} & \sci{8}{-3}   & \sci{3.7}{-3} & \Zn{*}        \\
                 &               &               & \Zn{16}       & \Zn{8.2}      & \Zn{8.0}      & \Zn{8.0}      &               \\ \hline
\end{tabular}}
\himasscaptionB{\mu}{|F|=2}
\label{tab-HMF2}
\end{table}
\end{center}
\newpage


\begin{table}[htb]
\begin{center}
\footnotesize{
\begin{tabular}{| c || c | c | c | c | c | c | c |} \hline
\multicolumn{4}{|c}{} & \multicolumn{4}{c|}{} \\
\multicolumn{1}{|c}{}&\multicolumn{2}{c}{\large{$e \to \tau$}} &\multicolumn{2}
{c}{\Large{ZEUS}} & \multicolumn{2}{c}{\large{$F=0$}} &\\ 
\multicolumn{4}{|c}{} & \multicolumn{4}{c|}{} \\ \hline
                 &               &               &               &               &               &               &               \\
\alphabeta       & \lbm{\ShL}    & \lbm{\ShR}    & \lbm{\tShL}   & \lbm{\VzL}    & \lbm{\VzR}    & \lbm{\tVzR}   & \lbm{\VoL}    \\ 
                 & \eua           & \euda          & \eda           & \eda           & \eda           & \eua           & \euuda         \\ \hline \hline
                 & \taupie       &  \taupie      & \taupie       &  $\mathbf G_F$        & \taupie       & \taupie       & $\mathbf G_F$         \\
 1 1             & $0.4$         & $0.2$         & $0.4$         & $0.2$         & $0.2$         & $0.2$         & $0.2$         \\
                 & \Zn{3.0}      & \Zn{2.5}      & \Zn{4.6}      & \Zn{3.3}      & \Zn{3.3}      & \Zn{2.4}      & \Zn{1.2}      \\ \hline 
                 &               & \tauKe        & \Kpinunu      & \tauKe        & \tauKe        &               & \Kpinunu      \\
 1 2             &               & $5$           & $10^{-3}$     & $3$           & $3$           &               & \sci{2.5}{-4} \\
                 & \Zbox{3.1}    & \Zbox{2.5}    & \Zn{4.7}      & \Zn{3.7}      & \Zn{3.7}      & \Zbox{2.7}    & \Zn{1.3}      \\ \hline
                 &               & \BtaueX       & \BtaueX       & \BlnuX        & \BtaueX       &               & \BlnuX        \\
 1 3             &  *            & $8$           & $8$           & $2$           & $4$           &  *            & $2$           \\
                 &               & \Zbox{5.1}    & \Zbox{5.1}    & \Zn{4.6}      & \Zn{4.6}      &               & \Zn{4.6}      \\ \hline
                 &               & \tauKe        & \Kpinunu      & \tauKe        & \tauKe        &               & \Kpinunu      \\
 2 1             &               & $5$           & $10^{-3}$     & $3$           & $3$           &               & \sci{2.5}{-4} \\
                 & \Zbox{16}     & \Zn{9.2}      & \Zn{12}       & \Zn{4.9}      & \Zn{4.9}      & \Zbox{6.2}    & \Zn{2.6}      \\ \hline
                 & \taueee       & \taueee       & \taueee       & \taueee       & \taueee       & \taueee       & \taueee       \\
 2 2             & $20$          & $30$          & $66$          & $33$          & $33$          & $10$          & $6.1$         \\  
                 & \Zbox{20}     & \Zbox{11}     & \Zbox{12}     & \Zbox{6.2}    & \Zbox{6.2}    & \Zn{11}     & \Zbox{4.3}    \\ \hline
                 &               & \BtaueX       & \BtaueX       & \BlnuX        & \BtaueX       &               & \BlnuX        \\
 2 3             & \Zn{*}        & $8$           & $8$           & $2$           & $4$           & \Zn{*}        & $2$           \\
                 &               & \Zn{16}       & \Zn{16}       & \Zn{12}       & \Zn{12}       &               & \Zn{12}       \\ \hline
                 &               & \BtaueX       & \BtaueX       &  $\mathbf V_{ub}$     & \BtaueX       &               & $\mathbf V_{ub}$      \\
 3 1             & \Zn{*}        & $8$           & $8$           & $0.2$         & $4$           & \Zn{*}        & $0.2$         \\
                 &               & \Zn{17}       & \Zn{17}       & \Zn{5.4}      & \Zn{5.4}      &               & \Zn{5.4}      \\ \hline
                 &               & \BtaueX       & \BtaueX       & \BlnuX        & \BtaueX       &               & \BlnuX        \\
 3 2             & \Zn{*}        & $8$           & $8$           & $2$           & $4$           & \Zn{*}        & $2$           \\
                 &               & \Zn{22}       & \Zn{22}       & \Zn{7.6}      & \Zn{7.6}      &               & \Zn{7.6}      \\ \hline
                 &               & \taueee       & \taueee       & \taueee       & \taueee       &               & \taueee       \\
 3 3             & \Zn{*}        & $30$          & $66$          & $33$          & $33$          & \Zn{*}        & $6.1$         \\
                 &               & \Zbox{30}     & \Zbox{30}     & \Zbox{15}     & \Zbox{15}     &               & \Zn{15}       \\ \hline
\end{tabular}}
\himasscaptionB{\tau}{F=0}
\label{tab-F0} 
\end{center}
\end{table}


\begin{table}[htb]
\begin{center}
\footnotesize{ 
\begin{tabular}{| c || c | c | c | c | c | c | c |} \hline
\multicolumn{4}{|c}{} & \multicolumn{4}{c|}{} \\
\multicolumn{1}{|c}{}&\multicolumn{2}{c}{\large{$e \to \tau$}} &\multicolumn{2}
{c}{\Large{ZEUS}} & \multicolumn{2}{c}{\large{$|F|=2$}} &\\ 
\multicolumn{4}{|c}{} & \multicolumn{4}{c|}{} \\ \hline
                 &               &               &               &               &               &               &               \\
\alphabeta       & \lbm{\SzL}    & \lbm{\SzR}    & \lbm{\tSzR}   & \lbm{\SoL}    & \lbm{\VhL}    & \lbm{\VhR}    & \lbm{\tVhL}   \\
                 & \euba          & \euba          & \edba          & \euddba        & \edba          & \eudba         & \euba          \\ \hline \hline
                 &  $\mathbf G_F$        & \taupie       & \taupie       & $\mathbf G_F$         & \taupie       & \taupie       & \taupie       \\
  1 1            & $0.3$         & $0.4$         & $0.4$         & $0.3$         & $0.2$         & $0.1$         & $0.2$         \\
                 & \Zn{5.4}      & \Zn{5.4}      & \Zn{7.1}      & \Zn{2.8}      & \Zn{2.6}      & \Zn{1.3}      & \Zn{1.7}      \\ \hline
                 & \Kpinunu      &               & \tauKe        & \Kpinunu      & \Kpinunu      & \tauKe        &               \\
  1 2            & $10^{-3}$     &               & $5$           & $10^{-3}$     & \sci{5}{-4}   & $3$           &               \\
                 & \Zn{14}       & \Zbox{14}     & \Zn{9.3}      & \Zn{4.6}      & \Zn{5.5}      & \Zn{4.5}      & \Zbox{8.2}    \\ \hline
                 & $\mathbf V_{ub}$      &               & \BtaueX       &  $\mathbf V_{ub}$     & \BtaueX       & \BtaueX       &               \\
  1 3            & $0.4$         &    *          & $8$           & $0.4$         & $4$           & $4$           & \Zn{*}        \\
                 &  *            &               & \Zn{12}       & \Zn{5.5}      & \Zn{8.4}      & \Zn{8.4}      &               \\ \hline
                 & \Kpinunu      &               & \tauKe        & \Kpinunu      & \Kpinunu      & \tauKe        &               \\
  2 1            & $10^{-3}$     &               & $5$           & $10^{-3}$     & \sci{5}{-4}   & $3$           &               \\
                 & \Zn{5.9}      & \Zbox{5.9}    & \Zn{7.8}      & \Zn{3.2}      & \Zn{2.5}      & \Zbox{1.3}    & \Zbox{1.6}    \\ \hline
                 & \taueee       & \taueee       & \taueee       & \taueee       & \taueee       & \taueee       & \taueee       \\
  2 2            & $20$          & $20$          & $66$          & $55$          & $33$          & $15$          & $10$          \\ 
                 & \Zbox{19}     & \Zbox{19}     & \Zbox{13}     & \Zbox{6.2}    & \Zbox{6.5}    & \Zbox{5.2}    & \Zbox{9.7}    \\ \hline
                 & \BlnuX        &               & \BtaueX       & \BlnuX        & \BtaueX       & \BtaueX       &               \\
  2 3            & $4$           & \Zn{*}        & $8$           & $4$           & $4$           & $4$           & \Zn{*}        \\
                 & \Zn{*}        &               & \Zn{17}       & \Zn{8.1}      & \Zn{11}       & \Zn{11}       &               \\ \hline
                 & \BlnuX        &               & \BtaueX       & \BlnuX        & \BtaueX       & \BtaueX       &               \\
  3 1            & $4$           & \Zn{*}        & $8$           & $4$           & $4$           & $4$           & \Zn{*}        \\
                 & \Zn{*}        &               & \Zn{9.3}      & \Zn{4.7}      & \Zbox{2.6}    & \Zbox{2.6}    &               \\ \hline
                 & \BlnuX        &               & \BtaueX       & \BlnuX        & \BtaueX       & \BtaueX       &               \\
  3 2            & $4$           & \Zn{*}        & $8$           & $4$           & $4$           & $4$           & \Zn{*}        \\
                 & \Zn{*}        &               & \Zn{21}       & \Zn{10.2}     & \Zn{7.6}      & \Zn{7.6}      &               \\ \hline
                 &               &               & \taueee       & \taueee       & \taueee       & \taueee       &               \\
  3 3            & \Zn{*}        & \Zn{*}        & $66$          & $55$          & $33$          & $15$          & \Zn{*}        \\
                 &               &               & \Zbox{30}     & \Zbox{16}     & \Zbox{15}     & \Zbox{15}     &               \\ \hline
\end{tabular}}
\himasscaptionB{\tau}{|F|=2}
\label{tab-F2} 
\end{center}
\end{table}


\clearpage

\clearpage


\clearpage

\newcommand{\limitcaption}[1]{
\caption{Upper limits on
  $\lambda_{e\qone}\sqrt{\br{#1 q}}$ vs. $\MLQ$ for a)
  scalar and b) vector LQs. The quark flavors that couple to the LQs
  in the initial state are shown in parentheses following the LQ species.
  Upper limits on $\lambda_{e\qone}$ under the assumption $\br{#1 q}=0.5$
  are shown in c) for scalar LQs and d) for vector LQs that couple
  to $d$-type quarks. Also shown are
  existing limits~\pcite{zfp:c61:613,pr:d62:055009,epj:c15:1}
 (dashed lines). The numbers in parentheses indicate
 the generations of the quarks that couple to the $e$ and the $#1$,
  respectively. The regions above the curves are excluded at the
  \CL{95}.}}


\begin{figure}[htb]
\epsfig{figure=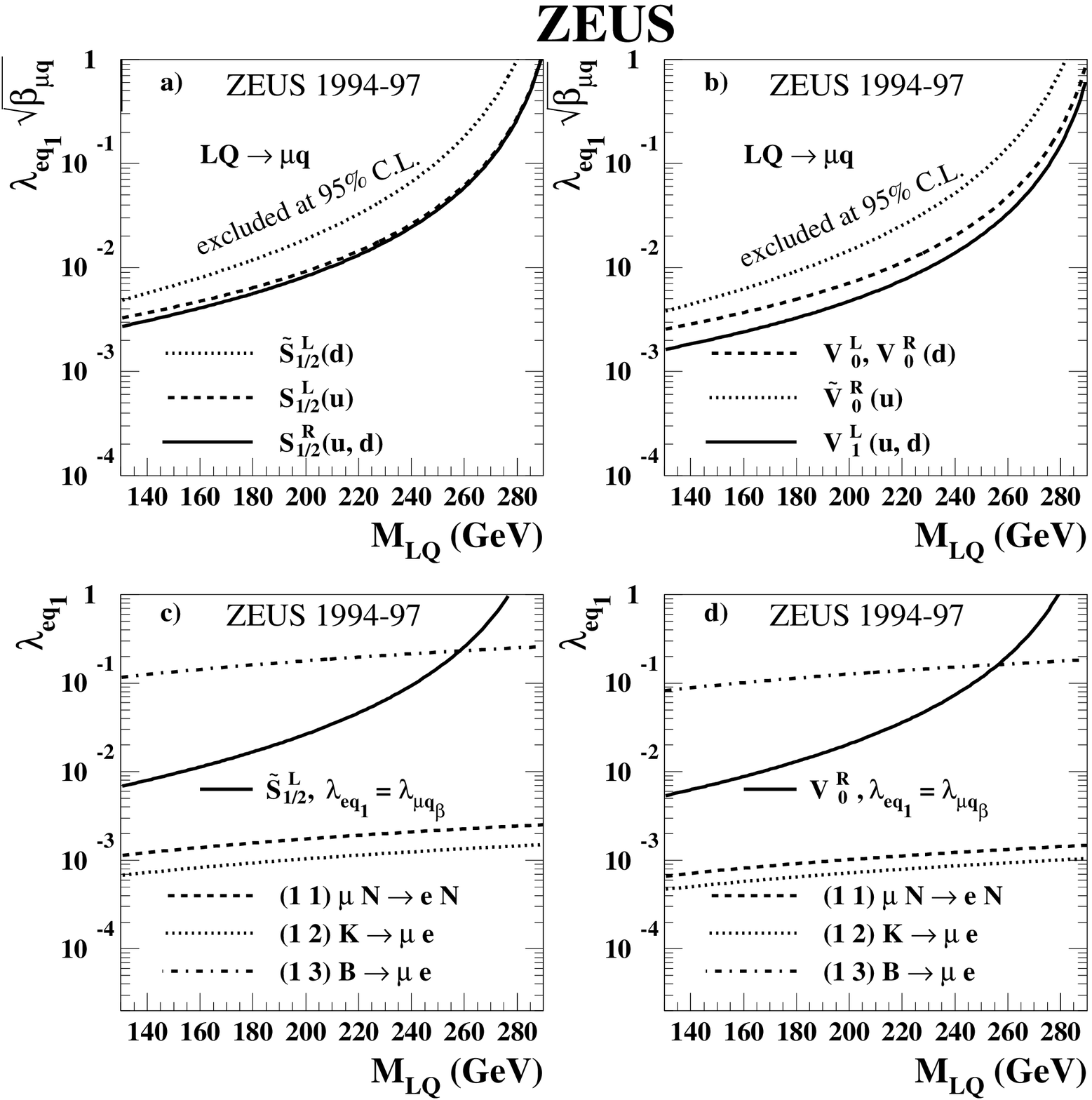,height=16cm}
\limitcaption{\mu}
\label{fig-mu_low}
\end{figure}
\clearpage


\begin{figure}[htb]
\epsfig{figure=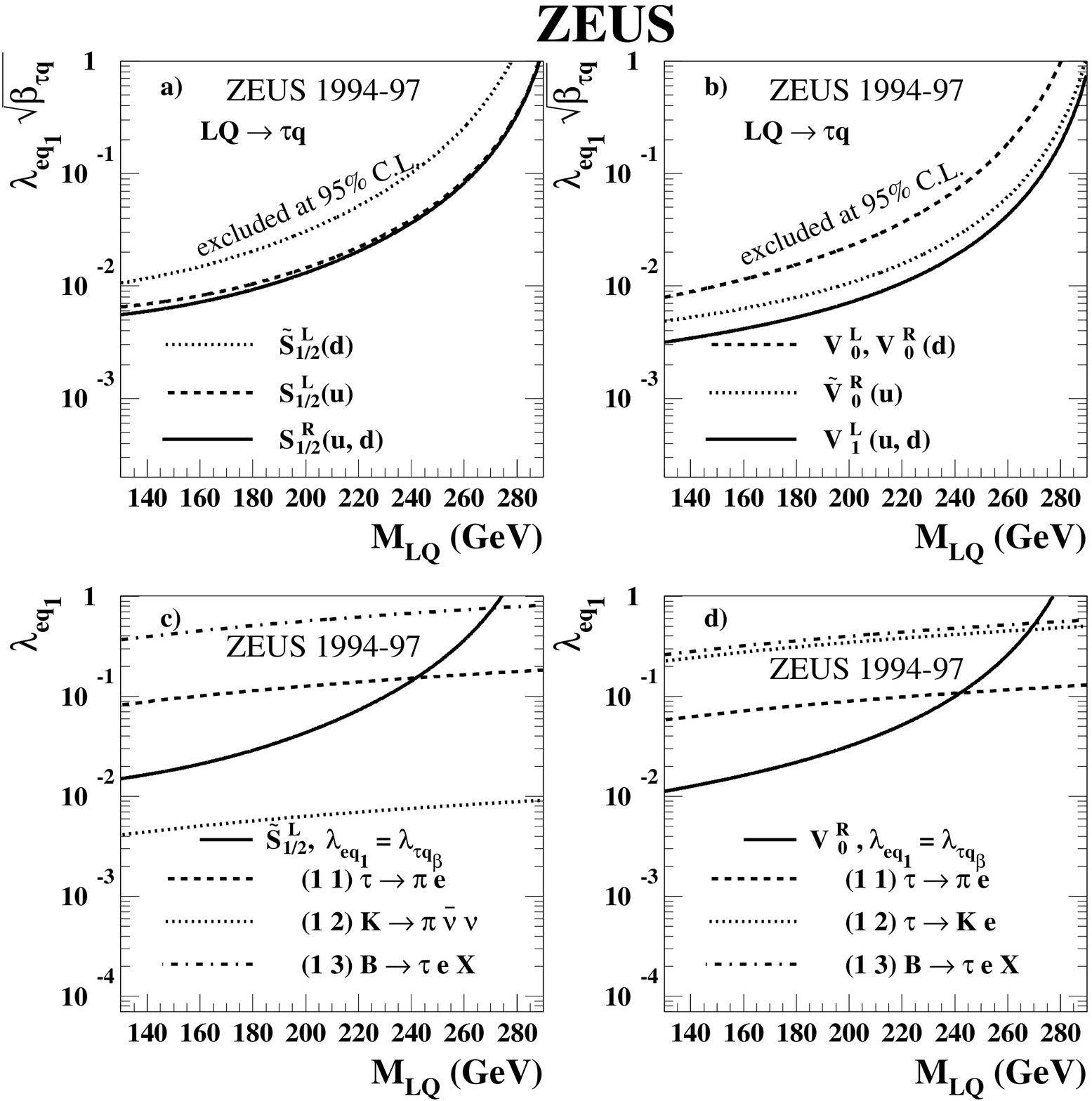,height=16cm}
\limitcaption{\tau}
\label{fig-tau_low}
\end{figure}


%
%
\end{document}